\documentclass[twocolumn]{aastex63_mod}

\usepackage{xspace}
\usepackage{amsmath}
\usepackage{hyperref}
\newcommand{\msun}{\hbox{$M_\odot$}\xspace}
\newcommand{\msyr}{\hbox{$M_\odot~\text{yr}^{-1}$}\xspace}

\newcommand{\flux}{\hbox{$\text{erg cm}^{-2} \text{ s}^{-1}$}\xspace}
\newcommand{\lum}{\hbox{$\text{erg s}^{-1}$}\xspace}
\newcommand{\um}{\hbox{$\mu$m}\xspace}
\newcommand{\Ha}{H$\alpha$\xspace}
\newcommand{\Hb}{H$\beta$\xspace}
\newcommand{\OIII}{[\ion{O}{3}] $\lambda 5007$\xspace}
\newcommand{\OIIIsimp}{[\ion{O}{3}]\xspace}

\newcommand{\OIIIl}{[\ion{O}{3}] $\lambda 4959$\xspace}

\newcommand{\NII}{[\ion{N}{2}] $\lambda 6584$\xspace}

\newcommand{\jhmag}{\hbox{$m_{J+JH+H}$}\xspace}

\newcommand{\mcsed}{\texttt{MCSED}\xspace}
\newcommand{\cloudy}{C{\footnotesize LOUDY}\xspace}
\usepackage{booktabs}
\usepackage{changepage}

\date{\today}
\shorttitle{Emission Line Galaxies at $1.2<z<1.9$}
\shortauthors{Nagaraj et al.}
\graphicspath{{./}{figures/}}

\begin{document}

\title{Measuring Stellar Masses of Emission Line Galaxies at $1.2<z<1.9$}

\correspondingauthor{Gautam Nagaraj}
\email{gxn75@psu.edu}

\author[0000-0002-0905-342X]{Gautam Nagaraj}
\affiliation{Department of Astronomy \& Astrophysics, The Pennsylvania
State University, University Park, PA 16802, USA}
\affiliation{Institute for Gravitation and the Cosmos, The Pennsylvania State University, University Park, PA 16802, USA}

\author[0000-0002-1328-0211]{Robin Ciardullo}
\affiliation{Department of Astronomy \& Astrophysics, The Pennsylvania
State University, University Park, PA 16802, USA}
\affiliation{Institute for Gravitation and the Cosmos, The Pennsylvania State University, University Park, PA 16802, USA}

\author{Alex Lawson}
\affiliation{Department of Astronomy \& Astrophysics, The Pennsylvania
State University, University Park, PA 16802, USA}

\author[0000-0003-4381-5245]{William P. Bowman}
\affiliation{Department of Astronomy \& Astrophysics, The Pennsylvania
State University, University Park, PA 16802, USA}
\affiliation{Institute for Gravitation and the Cosmos, The Pennsylvania State University, University Park, PA 16802, USA}

\author[0000-0003-2307-0629]{Greg Zeimann}
\affiliation{Hobby Eberly Telescope, University of Texas, Austin, TX 78712, USA}

\author[0000-0001-8835-7722]{Guang Yang}
\affiliation{Department of Physics and Astronomy, Texas A\&M University, College Station, TX 77843-4242, USA}
\affiliation{George P.~and Cynthia Woods Mitchell Institute for Fundamental Physics and Astronomy, Texas A\&M University, College Station, TX 77843, USA}

\author[0000-0001-6842-2371]{Caryl Gronwall}
\affiliation{Department of Astronomy \& Astrophysics, The Pennsylvania
State University, University Park, PA 16802, USA}
\affiliation{Institute for Gravitation and the Cosmos, The Pennsylvania State University, University Park, PA 16802, USA}

\begin{abstract}

The accurate measurement of stellar masses over a wide range of galaxy properties is essential for better constraining models of galaxy evolution. Emission line galaxies (ELGs) tend to have better redshift estimates than continuum-selected objects and have been shown to span a large range of physical properties, including stellar mass. Using data from the 3D-HST Treasury program, we construct a carefully vetted sample of 4350 ELGs at redshifts $1.16<z<1.90$. We combine the 3D-HST emission line fluxes with far-UV through near-IR photometry and use the \mcsed spectral energy distribution fitting code to constrain the galaxies' physical parameters, such as their star formation rate (SFRs) and stellar masses. Our sample is consistent with the $z\sim 2$ mass-metallicity relation. More importantly, we show there is a simple but tight correlation between stellar mass and absolute magnitude in a near-IR filter that will be particularly useful in quickly calculating accurate stellar masses for millions of galaxies in upcoming missions such as \textit{Euclid} and the \textit{Nancy Grace Roman Space Telescope}. 






\end{abstract}


\section{Introduction} \label{sec:intro}


Hydrodynamical and semi-analytic models of galaxy formation and evolution in the framework of the cold dark matter cosmology paradigm have advanced greatly over the past few decades, successfully reproducing many observable features in the universe. However the sheer complexity and multiple physical scales of the problem mean that we have a long way to go to achieve a complete understanding of the astrophysical processes that govern galactic formation \citep[e.g., see][and references therein]{SomervilleDave2015}. As models become more refined, it becomes necessary to provide increasingly precise and accurate observational constraints. One of the most important of these is stellar mass. 

Based on the strong correlations between stellar mass and other physical parameters such as metallicity, star formation rate (SFR), galaxy size, and morphology \citep[e.g.,][]{Gavazzi1996,GavazziScodeggio1996,Scodeggio2002,Kauffmann2003,Tremonti2004,Gallazzi2005,Noeske2007,Williams2010}, and the dependence of the stellar mass distribution on environment  \citep[e.g.,][]{Tomczak2017}, we can deduce that stellar mass must play an important role in regulating or affecting the formation and evolution of galaxies. Stellar mass has been shown to be closely related to the properties of satellite galaxies \citep{vandenBosch2008} and is tightly correlated with halo mass \citep{More2009,Yang2009,Leauthaud2012,Reddick2013,WatsonConroy2013,Tinker2013,Gu2016,Behroozi2019}. 


For these reasons, many efforts have been made to measure the stellar masses of galaxies throughout the course of the universe's history \citep[e.g.,][]{Marchesini2009,Ilbert2013,Moustakas2013,Muzzin2013uvj,Tomczak2014,Grazian2015,Song2016,Davidzon2017,Wright2018,Leja2020}. Typically, these studies have used broadband images in the rest-frame near-infrared (NIR) and/or ultraviolet (UV) to select their galaxy samples. Rest-frame red and NIR are a direct probe of the majority of stars in a galaxy and is minimally affected by ongoing star formation, which dominates the UV emission. Therefore, NIR selection is especially useful as a proxy for stellar mass. However, at higher redshifts the exposure times required to probe galaxies with low stellar mass (e.g., less than $10^9 \msun$) are formidable, especially with ground-based instruments \citep[e.g.,][]{Marchesini2009}. 


Galaxy surveys based on line emission have their limitations, often missing massive systems with high dust content and quiescent galaxies. However, they can identify objects that  continuum-selection cannot.  Moreover, unless color-selected galaxies are re-observed with long slit \citep[e.g.,][]{Steidel2004} or multi-slit  \citep[e.g.,][]{Davis2003,Lilly2007} spectrographs, emission-line selected objects will have more precise and accurate redshift measurements than studies based on photometry alone.  This is especially true if multiple emission lines can be identified.


High-redshift samples selected via emission lines themselves have traditionally been identified through the use of narrow-band filters, either in the optical or NIR \citep[e.g.,][]{Gronwall2007,Ouchi2008,Ciardullo2012,Sobral2013,Zheng2013,Suzuki2016,Ouchi2018,Shimakawa2018}. While such surveys are efficient and economical, their effective volume per observation is limited by the width of the filter's bandpass. In order to increase the observed redshift volume, slitless  \citep{Pirzkal2004,Pirzkal2013,Pirzkal2017,Brammer2012,Momcheva2016} or integral field unit \citep{Bacon2015, Hill2016} spectroscopy is needed.

Given the relatively few rest-frame UV emission lines in normal (non-active) galaxies, ELGs with redshifts 
$1\lesssim z \lesssim 2$ are difficult to detect from the ground.  For these galaxies, bright lines such as \OIII, \Hb, and \Ha fall in the NIR, where numerous atmospheric features severely restrict their observation. While instruments like MOSFIRE \citep{McLean2012} and KMOS \citep{Sharples2013} have revolutionized ground-based NIR spectroscopy, the advent of space-based instruments, such as the G141 grism on the WFC3 camera on the \textit{Hubble Space Telescope}, have greatly enhanced the identification and study of ELGs in this intermediate redshift regime.

The 3D-HST Treasury program (\citeauthor{Brammer2012} \citeyear{Brammer2012}; see \S \ref{sec:data} for details) has given rise to an extensive sample of high-redshift galaxies that are identified by their rest-frame optical emission lines on WFC3 G141 grism frames. By combining this grism data with the wealth of photometric data available for galaxies in the 3D-HST fields \citep[i.e.,][]{Skelton2014,Momcheva2016}, we can construct spectral energy distributions (SEDs) for a large sample of high-redshift galaxies. These data can then be fit using numerical models to infer physical properties for the galaxies, such as their SFRs, stellar masses, internal extinctions, and sizes. Using this rich data set, we can fully explore the complexities and intricacies of ELGs in a large swath of what was traditionally considered a redshift desert: $1.2<z<1.9$.

Stellar mass is typically measured using detailed SED modeling \citep[e.g.,][]{Walcher2011, Conroy2013}. SED fitting codes work by adopting a set of spectral templates that give the multiwavelength emission and absorption from stars, gas, dust, and AGN over a wide range of physical conditions, and then building complex stellar populations using some assumptions about an object's star formation history and dust attenuation \citep[e.g.,][and references therein]{Conroy2013}. By using a minimization algorithm (e.g., \texttt{MAGPHYS}, \citeauthor{daCunha2008} \citeyear{daCunha2008}; \texttt{FAST}, \citeauthor{Kriek2009} \citeyear{Kriek2009}) or Bayesian MCMC techniques (e.g., \texttt{GALMC}, \citeauthor{Acquaviva2011} \citeyear{Acquaviva2011}; \texttt{Prospector}, \citeauthor{Leja2017} \citeyear{Leja2017}; \mcsed, \citeauthor{Bowman2020} \citeyear{Bowman2020}), SED fitting codes attempt to converge on the best-fit physical parameters, including stellar mass.

In the coming decade,  \textit{Euclid} and the \textit{Nancy Grace Roman Space Telescope} (\textit{RST}) will create \OIIIsimp-detected slitless spectroscopic samples of millions of galaxies at redshifts $0.84<z<2.69$ and $1.00<z<2.85$, respectively, with corresponding NIR photometry. Fitting all these galaxies with state-of-the-art SED codes would likely be infeasible, but an expansion of the epoch's stellar mass database will greatly enhance our understanding of galactic evolution.  Here we use a sample of 4,300 \OIIIsimp-detected 3D-HST galaxies, well matched in redshift and emission line flux to the surveys of \textit{Euclid} and \textit{RST}, to model the behavior of stellar mass as a function of easily obtained observables.  We show that there is a tight relationship between stellar mass and absolute magnitude in a NIR filter. This correlation can be used to provide simple but accurate estimates of stellar mass for the \textit{Euclid} and \textit{RST} samples.


In \S \ref{sec:data}, we describe all of the data that have been used in our analysis of the properties of the dust, gas, and radiation at redshifts $1.2<z<1.9$. In \S \ref{sec:methods}, we discuss how we created a clean sample of ELGs from the original 3D-HST data (\ref{subsec:sample}), incorporated \textit{GALEX} and \textit{Swift} photometry into the sample (\ref{subsec:galexswift}), performed SED fitting (\ref{subsec:mcsed}), removed active galactic nuclei (AGN) from the sample (\ref{sec:AGN}), and created a companion photometric sample (\ref{sec:emvsphot}). In \S \ref{sec:individual}, we list the assumptions behind our SED fitting and show an example of the detailed results obtained for each individual galaxy and discuss the mass-metallicity correlation for our sample (\ref{sec:emprelmass}). In \S \ref{sec:massabsmag} we present the tight correlation between stellar mass and NIR absolute magnitude that can be used by future missions such as \textit{Euclid} and \textit{RST}. Finally, in \S \ref{sec:disc}, we summarize our investigation and look to the future. We assume a $\Lambda$CDM cosmology with $\Omega_{\Lambda} = 0.69$, $\Omega_M = 0.31$ and $H_0 = 69$~km~s$^{-1}$~Mpc$^{-1}$ \citep{Bennett2013}. All magnitudes given in the paper are in the AB magnitude system \citep{Oke1974}.


This paper is the first of three papers analyzing 3D-HST sources at redshifts $1.2<z<1.9$. The second paper (Nagaraj et al.~submitted), which we will hereafter refer to as Paper~II, focuses on the relationships among stars, gas, and dust in galaxies with available mid- and far-IR data. Finally, in Paper~III (Nagaraj et al.~in prep) will derive the \Ha and \OIII luminosity functions based on our clean ELG sample and present a measurement of the bias of the ELG population. This bias will be especially useful for preparing for the \textit{Euclid} and \textit{RST} era, as the ELGs found via the WFC3's G141 grism will be very similar to those which will be identified by the grisms of these future missions. In Paper~III, we will also present our full catalog.

\section{Data}\label{sec:data}

The 3D-HST Treasury program \citep[GO-11600, 12177, 12328;][]{Brammer2012, Momcheva2016} is an emission-line survey  with the \textit{Hubble Space Telescope's\/} Wide Field Camera 3 G141 grism ($1.08 < \lambda < 1.67 \um, R \sim 130$). In 124 two-orbit-depth pointings within 4 of the 5 Cosmic Assembly Near-infrared Deep Extragalactic Legacy Survey (CANDELS) fields (described below), the survey reached a monochromatic flux limit of $\sim 10^{-17}$ \flux. As the fifth CANDELS field, the Great Observatories Origins Deep Survey (GOODS) North, was observed previously by A Grism H-Alpha SpecTroscopic survey \citep[AGHAST,][]{Weiner2014} with similar flux limits, these data produced measurements over an 625 arcmin$^2$ area of sky. For the five fields, \cite{Momcheva2016} found that the wavelength-dependent line sensitivity of the survey can be expressed via

\begin{equation} \label{eq:linesens}
    1 \sigma = 8 \times 10^{-18}\left(\frac{G(\lambda)}{G(1.5~\mu {\rm m})}\right)^{-2}\left(\frac{R}{5~{\rm pix}}\right)~\flux
\end{equation}
where $G(\lambda)$ is the G141 grism's throughput curve versus wavelength and $R$ is an object's flux radius in pixels from a \texttt{SExtractor} image analysis. 


The 3D-HST grism observations were made in a large swath of the five CANDELS fields \citep{Grogin2011,Koekemoer2011}, where extensive, deep multiband photometry is available. Using point spread function (PSF) matching techniques, \cite{Skelton2014} carefully combined the photometry of 147 different ground-based and space-based imaging data sets covering the wavelength range between 0.3 and 8 \um.  Included in this database are 23 photometric data points in the the All-Wavelength Extended Groth Strip International Survey \citep[AEGIS,][]{Davis2007}, 44 measurements in the Cosmological Evolution Survey \citep[COSMOS,][]{Scoville2007}, 18 measurements in the Ultra Deep Survey \citep[UDS,][]{Lawrence2007}, and 22 and 40 data points in GOODS North and South \citep[GOODS-N and GOODS-S,][]{Giavalisco2004}, respectively.

In parallel, \citet{Momcheva2016} used stacked F125W+F140W+F160W \textit{Hubble} images (\jhmag) to extract and fit grism spectra for nearly 80,000 unique sources down to $\jhmag=26$ (with $\sim 23,000$ of these sources brighter than $\jhmag=24$). This spectrophotometric data set features low noise and good spatial resolution by using interlaced (rather than drizzled) pixels and reducing contamination from overlapping spectra with EAZY SED fitting \citep{Brammer2008}. 

In our study, we focus on objects in the CANDELS fields in the redshift range $1.16 < z < 1.90$. Our investigation serves as a complement to that of \cite{Bowman2019}, who detailed the properties of emission-line objects in the $1.90 < z < 2.35$ redshift range; taken together, these two analyses include all sources where \OIIIsimp is detectable by the G141 grism.  In Paper~II, we delve into the properties of the $\sim 16$\% of galaxies in our sample that also have detections in the mid-IR and/or far-IR.

In our redshift range, the shortest rest-frame wavelengths covered by the \cite{Skelton2014} photometry vary from $\sim 1750$ to $\sim 1300$ \AA\null. Thus, for our lowest redshift objects, the rest-frame FUV is not well represented, restricting our ability to study dust attenuation. To improve upon this situation, we used the \textit{GALEX} Deep Imaging Survey \citep[DIS;][]{Martin2005,Morrissey2007}, which provides observed-frame NUV (and FUV) photometry down to $m_{AB}\sim 25$ in portions of the CANDELS fields\footnote{Data from \textit{GALEX} Data Release 6/7 (\url{http://galex.stsci.edu/GR6/}) include DIS matches for sources in all fields.}, as well as the deep GOODS-S survey conducted by the Ultraviolet/Optical Telescope on board the Neil Gehrels Swift Observatory \citep{Hoversten2009}. 

\section{Methods}\label{sec:methods}
\subsection{Vetting the Sample}\label{subsec:sample}

\begin{figure*}[!ht]
    \centering
    \resizebox{\hsize}{!}{
    \includegraphics[width=0.20\textwidth]{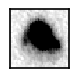}
    \includegraphics[width=0.8\textwidth]{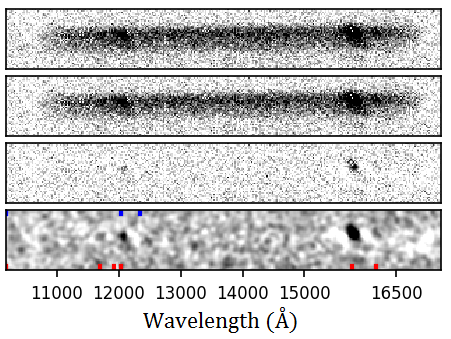}} 
    \resizebox{\hsize}{!}{
    \includegraphics{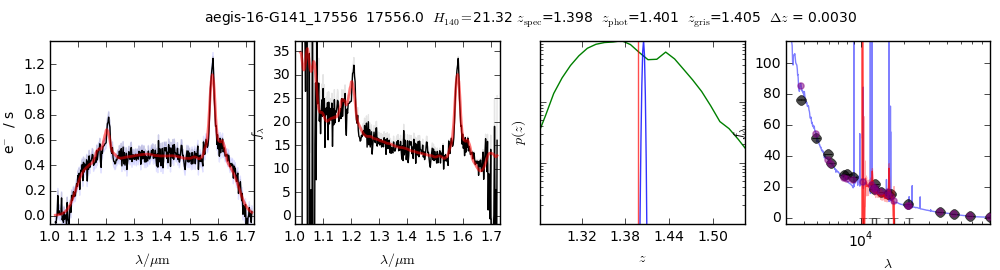}}
    \caption{Information used to rate the 3D-HST spectrum of the emission-line galaxy AEGIS 17556.  The top left shows the object's appearance in the WFC3 F140W filter; the top right shows the object's 2D-grism spectra in various stages of analysis (from top to bottom:  reduced, contamination-subtracted, continuum-subtracted, and smoothed with a Gaussian kernel of $\sigma = 1.5$ pixels).  The red tick marks indicate the expected locations of (from left to right) \Hb, [\ion{O}{3}]  $\lambda 4959,5007$, \Ha+\NII, and [\ion{S}{2}] $\lambda\lambda 6716,6731$.  The blue tick marks represent the observed locations of emission lines if the line identified as \OIIIsimp is actually \Ha.  The two left-most images on the bottom row show the object's 1-D spectrum (in electron counts and flux in units of $10^{-17}$ \flux). Third image on the bottom shows the probability distribution of the redshift determination, with the green line representing the photometric redshift, and the blue line showing the distribution with the inclusion of the grism information.  The final image on the bottom shows the galaxy's SED, including best-fit curve using the grism redshift. Based on the figures above, this source made it into our ELG sample.}
    \label{fig:rating}
\end{figure*}

With the relatively coarse resolution of the G141 grism, contamination by bright objects dispersed into the grism frame can lead to incorrect or spurious redshifts and emission line strengths. While \cite{Momcheva2016} designed a comprehensive algorithm to model and remove contamination from overlapping sources, failures of the program are inevitable.  Since the 3D-HST team's visual  inspection only extended down to $\jhmag=24$, additional vetting is needed to remove interlopers from the vast majority of fainter galaxies.  To create this sample, we followed a process similar to that described by \cite{Zeimann2014} and expanded upon by \cite{Bowman2019}. We first selected all $1.16 < z < 1.90$ sources in the 3D-HST catalog with well-determined redshifts derived from combining grism data and photometry, i.e., with a 68\% redshift confidence intervals less than $\Delta z = 0.05$.   We then created a webpage\footnote{\url{https://github.com/grzeimann/DetectWebpage/}} for each object containing the following information: the grism ID number, equatorial coordinates, and JH magnitude (\jhmag), F140W \textit{Hubble}/WFC3 image, the 2D grism image for four different stages of reduction (reduced, contamination-subtracted, continuum-subtracted, and smoothed with a 2D Gaussian kernel of $\sigma=1.5$ pixels), the 1D grism spectrum, the redshift probability distribution (including the photometric as well as the photometric+grism-based estimate), and the best-fit SED along with photometry from \cite{Skelton2014}. Figure~\ref{fig:rating} shows an example of the information for an AEGIS field galaxy. 

Table \ref{tab:source_counts} gives the number of sources in each field used for analysis. Fewer than half of the initial candidates made it into our (clean) ELG sample (first vs.\ second column), which is reasonable given all of the issues of contamination in a grism survey, especially for very faint sources. Around half of all ELGs in our sample are fainter than JH magnitude 24 (third column), meaning that pushing down to JH magnitude 26 nearly doubled the sample size of vetted objects.

Based on a visual inspection of these data, we assigned a set of indices to each object, in order to quantify our trust in the measurements. Specifically, we created five quality indices: 1) overall trust in the redshift measurement (based on all aspects of the diagnostic), 2) trust in or evidence for correctly identified emission lines, 3) whether or not there are missing sections of grism data in wavelength space (from, for example, a source appearing at the edge of a frame), 4) the existence of continuum-like contamination (from another source or because of a poorly modeled continuum), and 5) the existence of line-like contamination (from emission lines of other sources).

The redshift trust quality index was set to an integer value between $-1$ and 2; a score of 2 signifies our confidence that the redshift calculation is correct. The emission line index trust index ranged between $-1$ and 3 with scores of 2 to 3 indicating strong evidence for emission lines. The three remaining flags were set to values between 0 and 2, with 0 indicating the cleanest spectra in each case. Objects with a redshift trust scores of 2, emission line scores of 2 or 3, and contamination scores of 0 or 1 (little to no contamination) were included in our sample.

All heavily contaminated objects and moderately contaminated sources with poor models for the contamination were removed. The majority of contaminants are local objects whose distributions are unrelated to that of our sources. However, for a small fraction of sources ($\sim 10\%$), the contaminants are objects within the redshift range itself. While the censoring process slightly affects the clustering properties of our sample, it has negligible effects on the results presented in this paper. We will delve further into this issue with respect to the measurement of bias in Paper~III.


In order to characterize the accuracy of the grism redshifts, we compared the grism-$z$'s to ground-based spectroscopic measurements for the 281 objects with such measurements. These sources are listed in \cite{Skelton2014}.  Whenever possible, they took only the spectroscopic redshifts with the best confidence flag values. We use the normalized median absolute deviation to describe the spread of redshift error in the grism-$z$ sample, 

\begin{equation} \label{eq:sigmanmad}
    \sigma_{\rm{NMAD}}=1.48 \times {\rm{median}}\left( \frac{|\Delta z - {\rm{median}}(\Delta z)|}{1+z_{\rm{spec}}} \right)
\end{equation}
where $\Delta z = z_{\rm{grism}} - z_{\rm{spec}}$. For the sample, we find $\sigma_{\rm{NMAD}}=0.0017$. In addition, if we adopt the outlier definition from \cite{Momcheva2016} of $|\Delta z|/ 1 + z_{\rm spec} > 0.1$, we find an outlier fraction of 1.4\%. Such low overall error and outlier fraction is a testament to the high accuracy and precision of the grism redshifts.

We mention the caveat that the sources with spectroscopic redshifts span a narrower range of physical properties than the overall ELG sample, with a median JH magnitude of 22.7 compared to 24.0 and minimum of 24.5 vs 26.0. While it is possible that the fainter ELGs have a higher average error, our strict vetting process should limit such effects. Furthermore, in the range of magnitudes covered by the spectroscopic redshifts, we find no correlation between error and magnitude.

In terms of the emission lines used in the vetting process, the distinctively shaped blend of [\ion{O}{3}] $\lambda\lambda 4959,5007$ falls within the G141 grism throughout our entire redshift range of $1.16<z<1.90$\footnote{While \OIII does indeed occupy the entire range, the G141 throughput is very low shortward of 1.13$~\um$.  This limits object detections at redshifts $z < 1.25$.}, and \Ha is very prominent in most sources with $z<1.54$. The detection of these two features determines whether or not a galaxy enters our ELG sample.

\begin{figure*}
    \centering
    \includegraphics{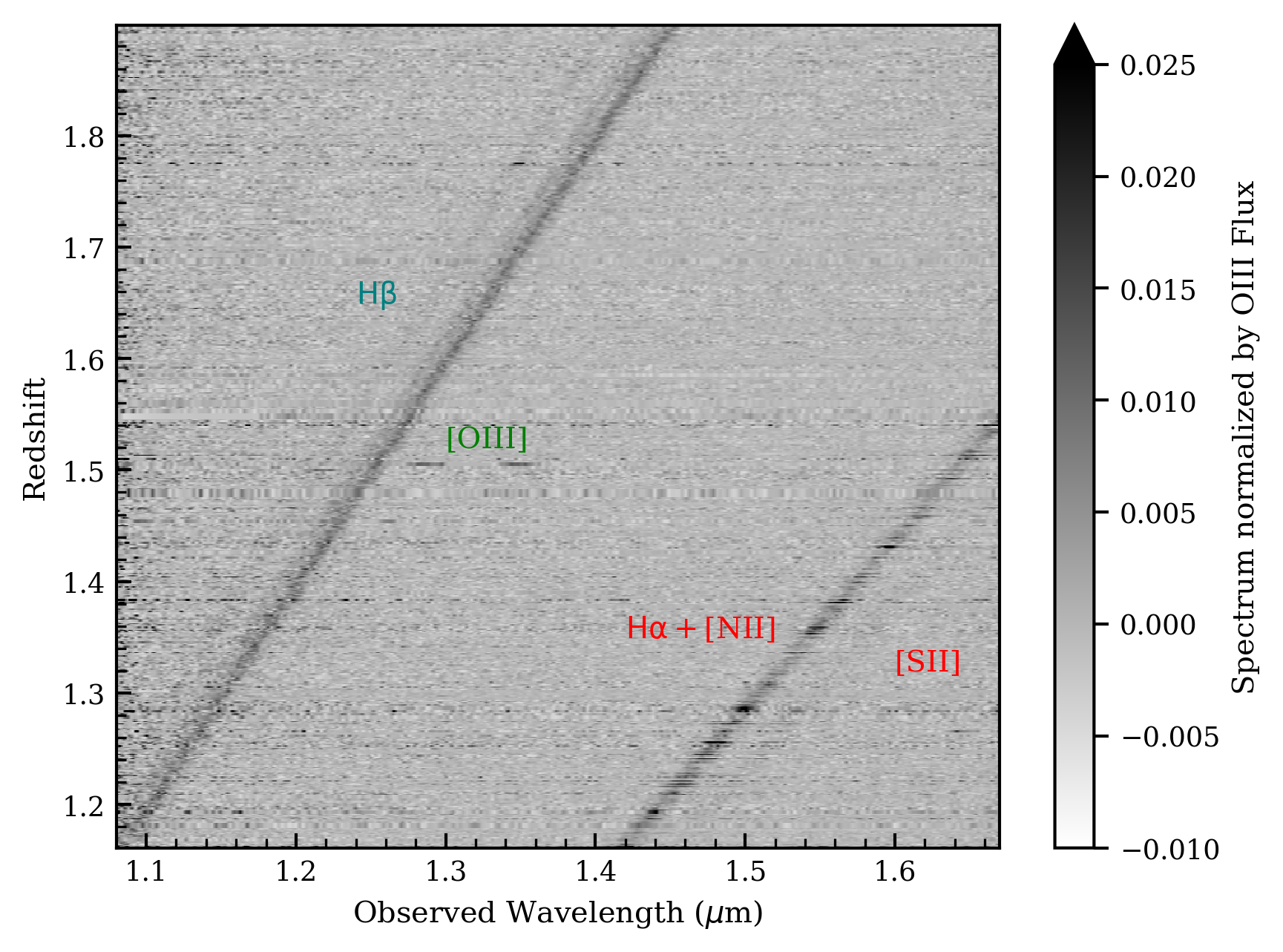}
    \caption{Continuum-subtracted 1-D spectra for 1,344 3D-HST ELGs at redshifts $1.16<z<1.9$, normalized by their \OIII flux. The lines \Ha, \Hb, and [\ion{O}{3}] $\lambda \lambda~4959,~5007$ are clearly visible, with \OIIIsimp occupying the entire redshift range.}
    \label{fig:ELSSpec}
\end{figure*}

Additional lines in the grism spectra may include \Hb ($z>1.22$) and occasionally the blended [\ion{S}{2}] doublet $\lambda \lambda 6717,6732$ ($z<1.48$). Figure \ref{fig:ELSSpec} shows the continuum-subtracted 1-D spectra for 1,344 vetted sources, with extreme negative values set to zero. The emission lines from \Ha, \Hb, and [\ion{O}{3}] $\lambda \lambda~4959,5007$ are clearly visible. As stated previously, the \OIIIsimp doublet is particularly useful for redshift determination given the telltale asymmetric shape from the relative strengths of \OIII and \OIIIl. 

For our sample, the observed fluxes of \Ha and \OIII are similar.  In the redshift range where both lines are within the grism coverage, 85.6\% of the sources feature both lines;  for the remaining galaxies, \Ha is visible in 73\% and \OIII in 27\%.  At least one of the these two lines is seen in every galaxy in our sample.

\begin{figure*}
    \centering
    \resizebox{\hsize}{!}{
    \includegraphics{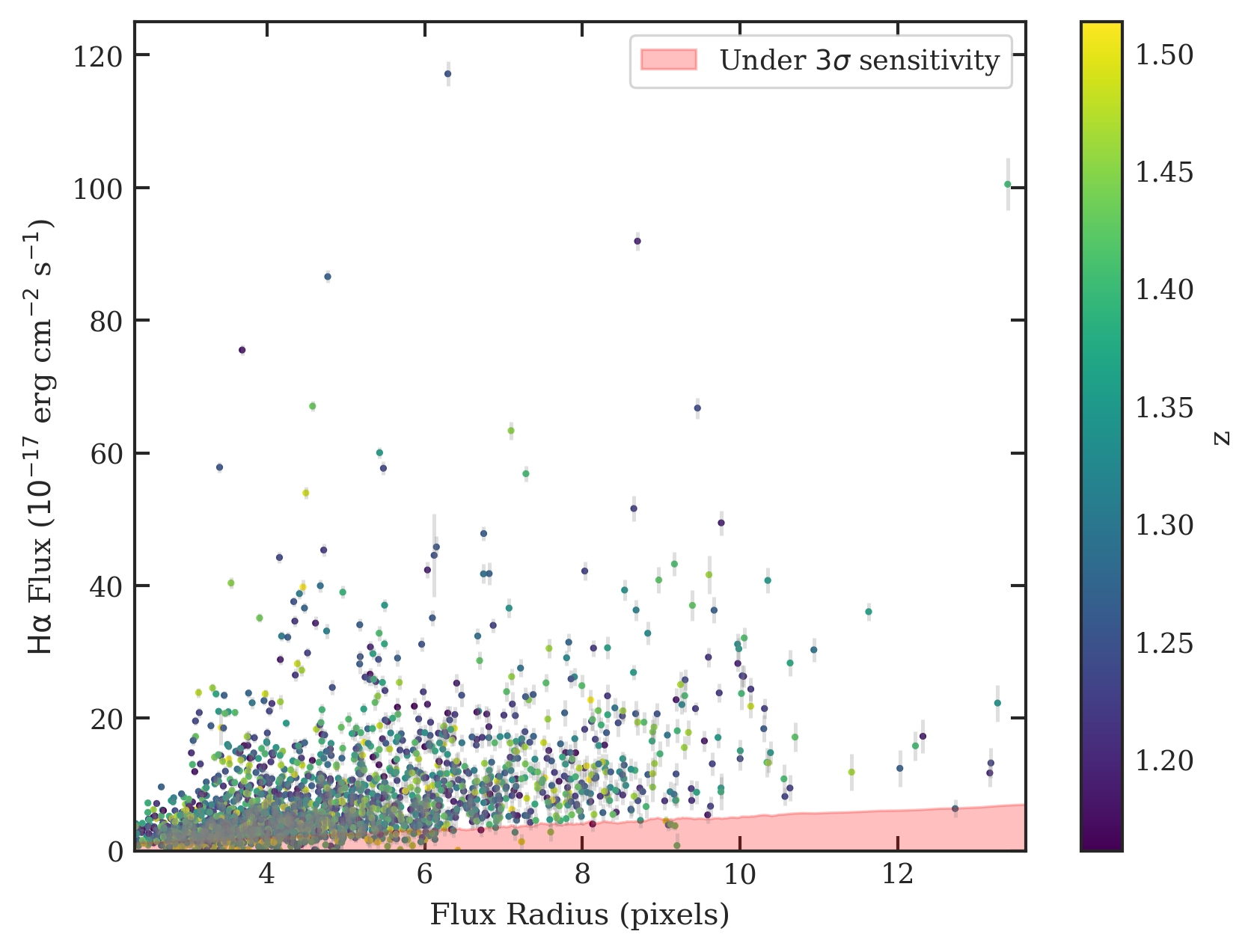}
    \includegraphics{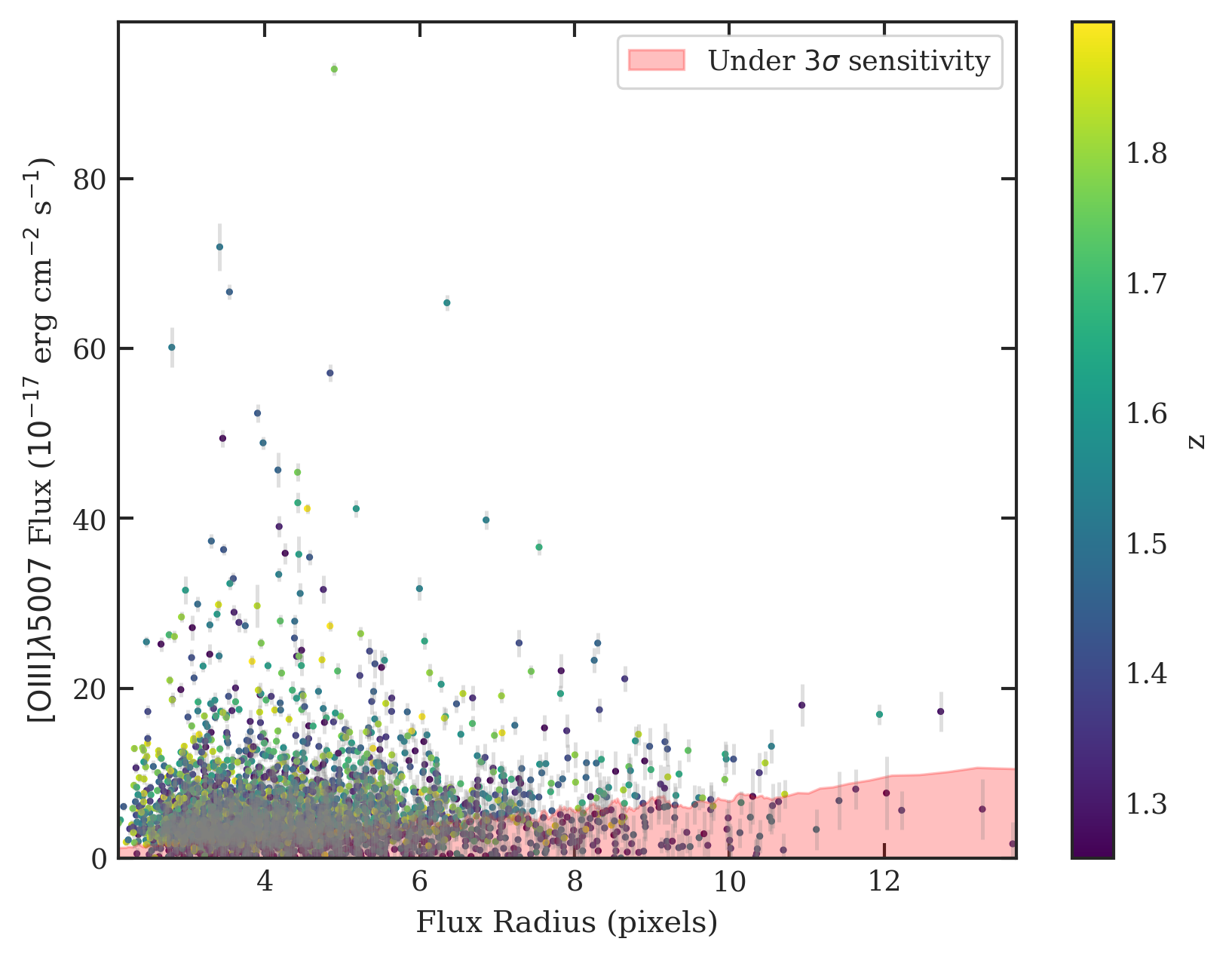}
    \includegraphics{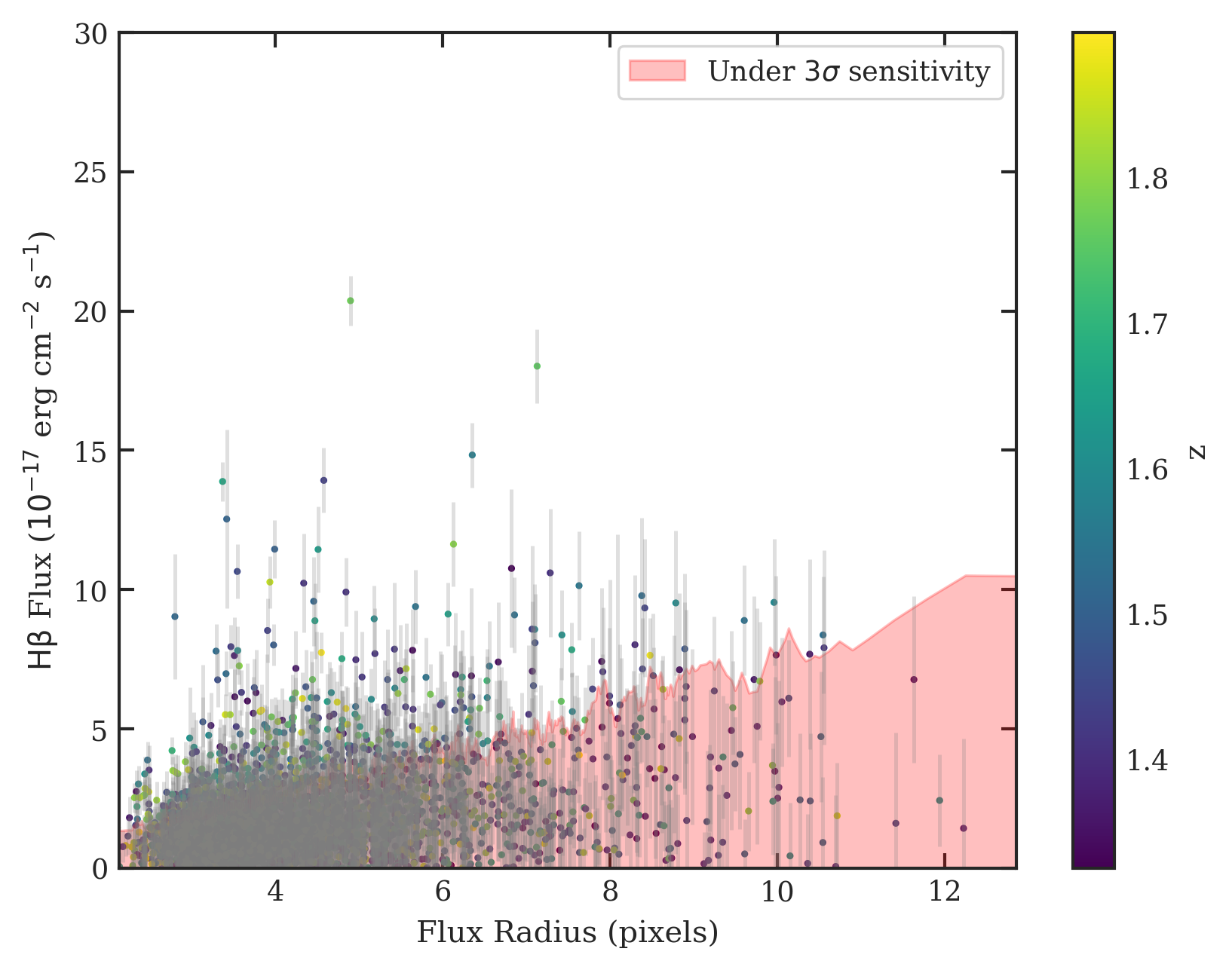}}
    \caption{\Ha (left), \OIIIsimp (middle), and \Hb (right) fluxes versus \texttt{SExtractor Flux\_Radius} \citep{Momcheva2016}. Regions where fluxes are under the $3 \sigma$ sensitivity limit, as determined by Equation \ref{eq:linesens}, are shaded in red (with slight smoothing). While the vast majority of \Ha and \OIIIsimp fluxes are above the sensitivity limit, most of the \Hb measurements and a non-negligible fraction of \Ha and \OIIIsimp fluxes lie within the shaded region, suggesting caution when directly applying the line ratios to problems such as nebular reddening. }
    \label{fig:HaOIIIcomplete}
\end{figure*}

Next, in Figure \ref{fig:HaOIIIcomplete} we show the distribution of \Ha, \OIIIsimp, and \Hb fluxes as a function of source size \citep[as given by the half-light \texttt{SExtractor Flux\_Radius;}][]{Momcheva2016}, with regions under $3\,\sigma$ sensitivity (calculated using Equation \ref{eq:linesens}) shaded in red. While most \Ha and \OIIIsimp fluxes are over the sensitivity limit, the majority of \Hb fluxes and a non-negligible number of \Ha and \OIIIsimp fluxes fall within the shaded region, suggesting possible incompleteness in this range. 

\begin{figure*}
    \centering
    \resizebox{\hsize}{!}{
    \includegraphics{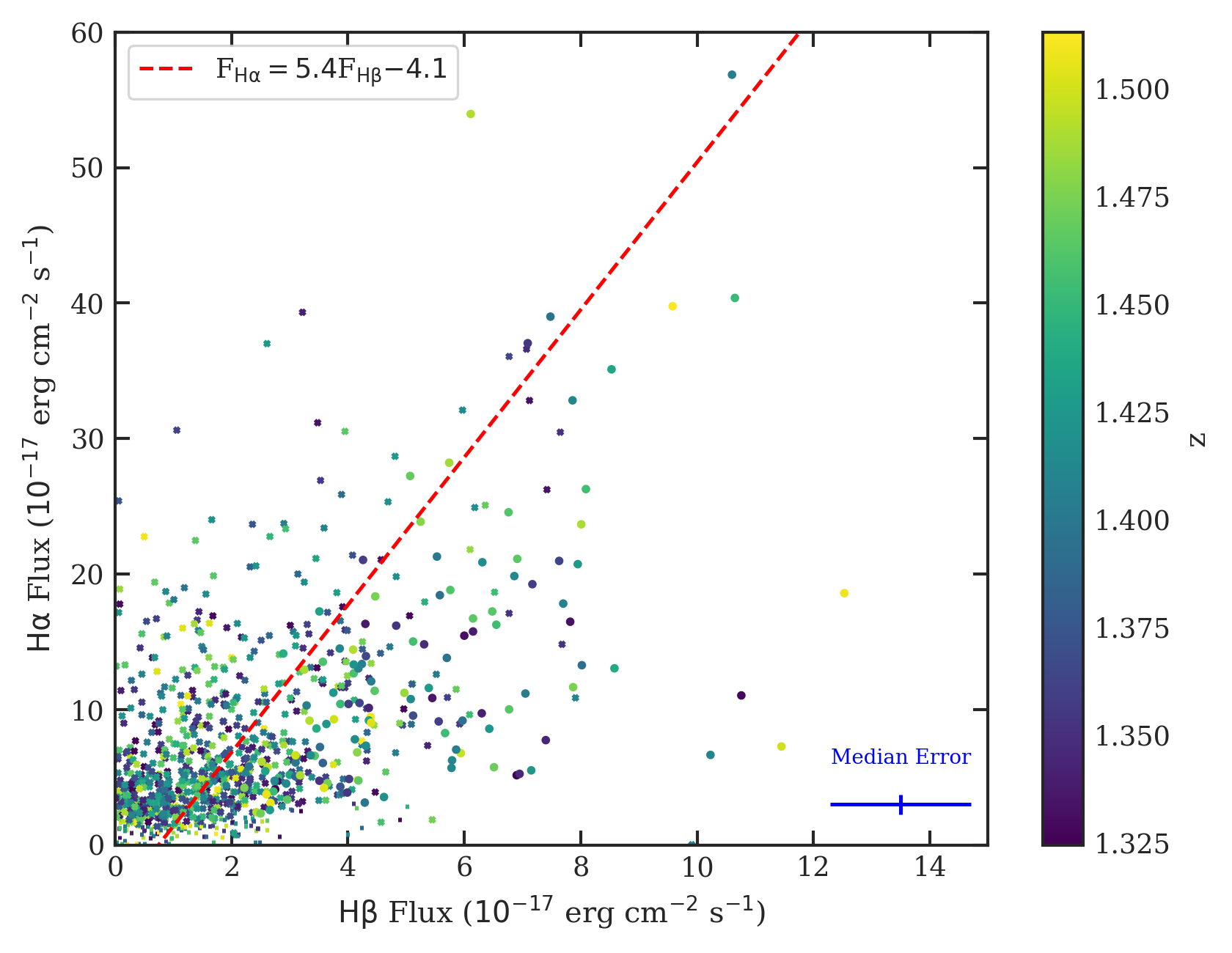}
    \includegraphics{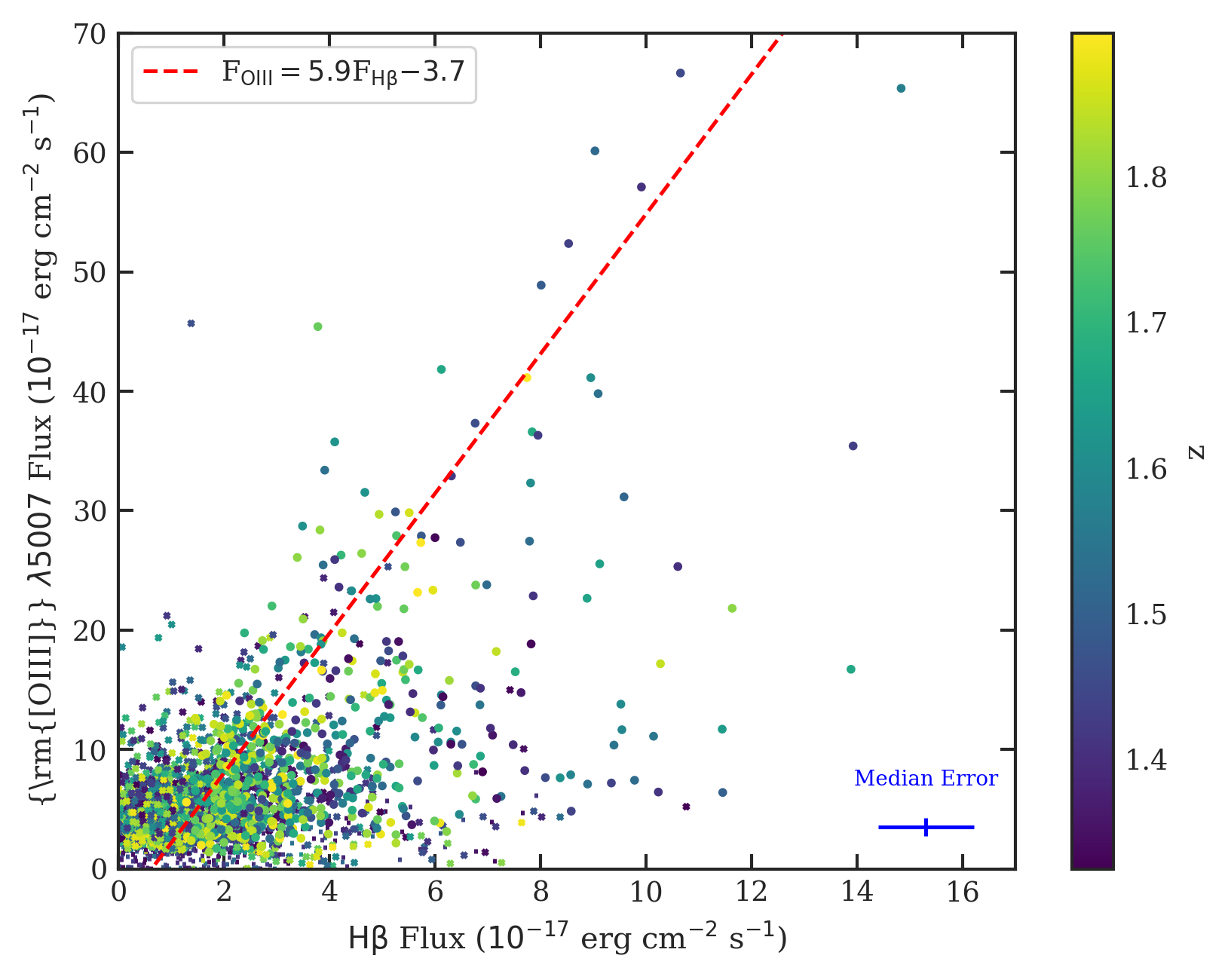}}
    \caption{\Ha versus \Hb fluxes (left) and \OIII versus \Hb fluxes (right) as measured by \cite{Momcheva2016}. A best fit line has been made using orthogonal distance regression. The largest points (circles) have both fluxes above the $3 \sigma$ sensitivity limit given by Equation \ref{eq:linesens}. The medium-sized ``x''-marks have one flux above its limit. Finally, the small vertical lines have neither flux above the limit. The median \Ha/\Hb ratio of $2.4$ suggests the absence of dust, whereas the ODR best-fit slope of $5.4$ implies moderately strong internal reddening. The truth is likely somewhere in the middle. \OIIIsimp tends to be pretty strong, with values comparable to those of \Ha. This justifies our assumption of using $\log U = -2.5$ for the ionization parameter in SED fitting.}
    \label{fig:HaHb}
\end{figure*}

The often poor \Hb measurements lead to unreasonably high and low \Ha/\Hb and \OIIIsimp/\Hb ratios in a large number of objects.  We show this in  Figure~\ref{fig:HaHb} where we plot \Ha vs.~\Hb and \OIIIsimp vs.~\Hb, along with the best-fit orthogonal distance regression (ODR) line. The circles (largest markers) are for sources with both flux measurements over their sensitivity limits. The ``x'' marks (medium-sized markers) are for sources with one flux measurement over its sensitivity limit. Finally, the small vertical lines are for sources where neither flux measurement is above the sensitivity limit. The points farthest away from the best-fit relations tend to be the small vertical lines (with both fluxes under the sensitivity limit), suggesting that we cannot rely on individual measurements in this regime.


The median \Ha/\Hb value for sources where both fluxes are over their sensitivity limits is $2.4$. Given that the physics of the Balmer decrement demands that the intrinsic H$\alpha$/H$\beta$ ratio be within $\sim 10\%$ of $2.86$ \citep{OsterbrockFerland2006}, this would seem to suggest an absence of dust in our sample of ELGs. On the other hand, as shown in Figure \ref{fig:HaOIIIcomplete}, the ODR of \Ha vs.~\Hb has a slope of $5.4$ ($6.2$ if we consider only those objects with both lines above the sensitivity limit).  This measurement implies the existence of a significant amount of internal reddening.  The truth is likely somewhere in between, due to the large scatter in values and potential errors in either \Ha or \Hb fluxes. We discuss nebular extinction in much greater detail in Paper~II.

The median \OIII/\Hb ratio for objects with line fluxes over our sensitivity limits is $2.6$, and the slope of the ODR fit is $5.9$ ($6.9$ above the sensitivity limits). In either case, the strong \OIIIsimp emission suggests a high ionization parameter, as we discuss in \S \ref{sec:individual}.

\begin{figure}
    \centering
    \resizebox{\hsize}{!}{
    \includegraphics{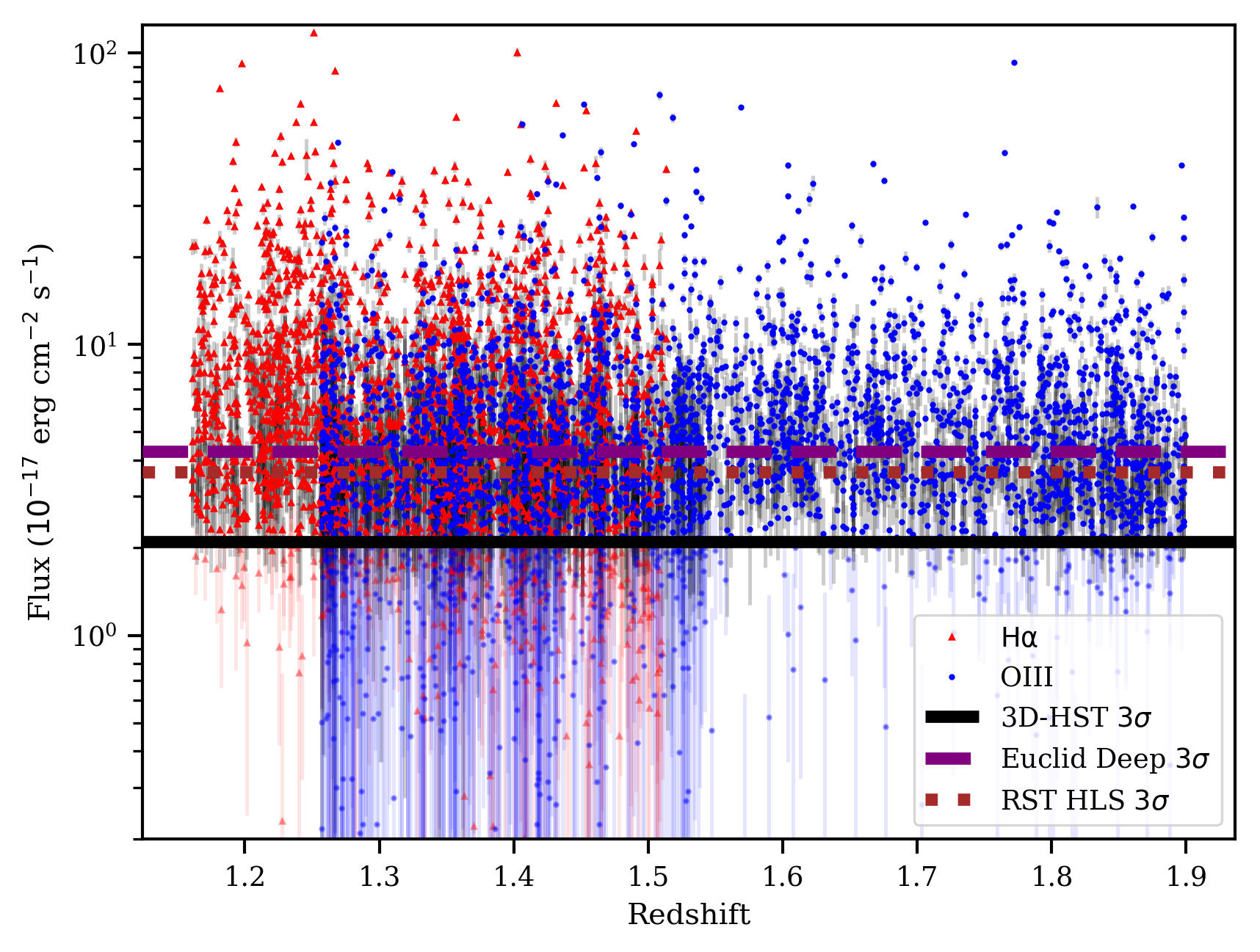}}
    \caption{\Ha and \OIII fluxes as a function of redshift, along with 
    simple $3 \sigma$ detection limits for 3D-HST, the \textit{Euclid} Deep Survey, and the \textit{RST} High Latitude Survey.}
    \label{fig:EuclidRST}
\end{figure}

Looking forward to \textit{Euclid} and \textit{RST}, in Figure \ref{fig:EuclidRST} we plot the 3D-HST \Ha and \OIII fluxes vs.~redshift with rough $3 \sigma$ detection limits (measured or predicted) for the 3D-HST survey, the \textit{Euclid} Deep Survey \citep{Laureijs2011}, and the \textit{RST} High Latitude Survey \citep{Spergel2015}. While the \textit{Euclid} and \textit{RST} surveys will have slightly brighter detection limits than 3D-HST, they will provide much larger galaxy samples given the much larger survey volumes.

\begin{table}[]
\begin{adjustwidth}{-1.8cm}{}
\begin{tabular}{@{}lcccc@{}}
\toprule
Field & Candidates & \# of ELGs & $m>24$ \\ \midrule
AEGIS & 2000 & 1196 & 593 \\
COSMOS & 1869 & 898 & 468 \\
GOODS-N & 1599 & 558 & 278 \\
GOODS-S & 2066 & 655 & 324 \\
UDS & 1807 & 1043 & 511 \\ \midrule
Total & 9341 & 4350 & 2174 \\ \bottomrule
\end{tabular}
\end{adjustwidth}
\caption{Number of sources in our sample. The first column gives the field name. The second column lists the initial number of $1.2<z<1.9$ candidates with a 68\% redshift confidence interval of $\Delta z < 0.05$  The third column represents the number of sources we consider to have well-measured optical emission lines. The last column gives the number of these sources fainter than magnitude 24. }
\label{tab:source_counts}
\end{table}

\subsection{Incorporation of \textit{GALEX} and \textit{Swift} Data} \label{subsec:galexswift}

Given that the \citet{Skelton2014} photometric catalog samples only rest-frame wavelengths down to $\sim 1400$~\AA, we used data from the \textit{GALEX} Deep Imaging Survey \citep[DIS;][]{Martin2005,Morrissey2007} and the \textit{Swift} GOODS-S survey \citep{Hoversten2009} to better constrain the rest-frame FUV spectra of our 3D-HST sample. Here, we describe the details of incorporating these data.

The issue with using \textit{GALEX} data is that the $5\farcs3$ full width half maximum (FWHM) of the instrument's point spread function (PSF) is $\sim 30$ times larger than that of the WFC3's NIR frames, making source counterpart identification difficult.  To address this problem, for every $z\leq 1.5$ 3D-HST galaxy in our sample, where the \textit{GALEX} NUV bandpass is shortward of the Lyman break ($912$~\AA), we searched for the nearest \textit{GALEX} source within a specified radius. 

\begin{figure}
    \centering
    \resizebox{\hsize}{!}{
    \includegraphics{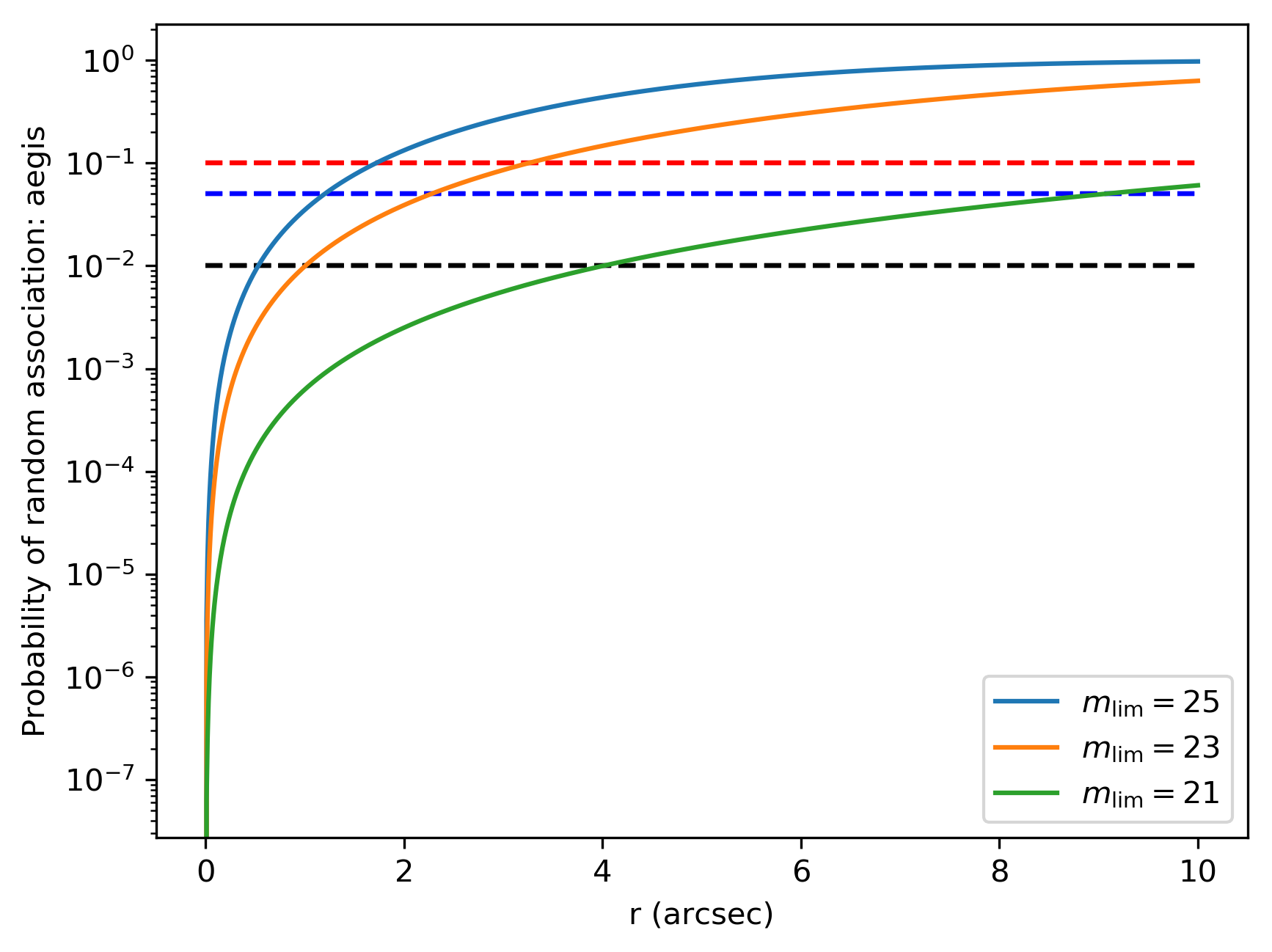}}
    \caption{Probability of a random association between a 3D-HST source in AEGIS and a \textit{GALEX} source with a NUV magnitude brighter than $m_{NUV} = 25$, 23, and 21. The dashed lines indicate confidence levels of 90\% (red), 95\% (blue), and 99\% (black).}
    \label{fig:GalexContam}
\end{figure}

In order to determine the aforementioned specified radius, we designed an experiment to evaluate the probability that a given match is random. The steps are described below.
\begin{enumerate}
    \item To mitigate the problem of single objects having multiple listings in the \textit{GALEX} catalog, we performed an average-distance hierarchical clustering calculation down to a separation of $0\farcs 53$ (1/10 the FWHM \textit{GALEX} NUV filter's PSF) to determine the actual number of sources within a given radius of each field's center.  This experiment showed that the density of \textit{GALEX} DIS sources is relatively uniform within 0.2$^{\circ}$ of the center of each CANDELS field.
    \item For each field, we obtained a list of all \textit{GALEX} sources within a distance $r$ of a 3D-HST galaxy located within $d = 0.1^{\circ}$ of the field center.  We did this using \textit{GALEX} subsamples of objects with NUV magnitudes $\leq 25$, $\leq 23$, and $\leq 21$. 
    \item Given the uniformity of the \textit{GALEX} sources, we calculated the probability of a given match to a 3D-HST source being random as $P=1-\left[1 - (r/d)^2 \right]^N$, where $N$ is the number of sources found in step 2.
    \item For each field, we determined the radius, as a function of galaxy magnitude, within which the probability of a random association was under $5\%$, and used this as a cutoff distance for accepting \textit{GALEX} matches.  These relations are shown in Figure \ref{fig:GalexContam}.
\end{enumerate}

Using the distribution of 3D-HST/\textit{GALEX} separations $f(r)$, the probability of random association $P(r)$, and the $r_{\rm max}$ values determined from the steps outlined above, we then calculated the mean probability of a \textit{GALEX} measurement being spurious as
\begin{equation} \label{eq:meanp}
    \langle P \rangle = \frac{\int_0^{r_{\rm max}}P(r)f(r)dr}{\int_0^{r_{\rm max}}f(r)dr}
\end{equation}
Based on the nearly uniform distance distribution, we expect only 9 of the accepted 517 \textit{GALEX} matches to be coincidences.

The above analysis shows that most of our 3D-HST - \textit{GALEX} associations are likely to be correct. However, the \textit{GALEX} measurements may still be affected by confusion: given the much larger PSF ($5\farcs3$ FWHM) of the data, each NUV source may contain contributions from several galaxies. 

In order to at least partially correct for confusion, we recorded the observed-frame U-band ($\lambda_{\rm eff} \sim 3800$~\AA) fluxes for all $z<=1.5$ objects within a \textit{GALEX} FWHM ($5\farcs3$) of our 3D-HST coordinates. The redshift cut was applied to avoid sources with nonzero U-band fluxes but near-zero \textit{GALEX} fluxes due to the differing rest-frame coverage of the two bands, specifically, the presence of the Lyman limit in latter band.  We then assumed that the fraction of the association's total NUV flux arising our 3D-HST galaxy was equal to that found for the U-band.  For those objects in our sample where the fraction of \textit{GALEX} NUV flux was less than 5\% the total flux for their association, we simply ignored the NUV measurement.  That reduced the sample of objects with \textit{GALEX} data from 517 to 394.

Of course, this correction assumes a linear correspondence between the observed-frame \textit{GALEX} NUV flux and that for the U-band.  This is almost certainly not true, especially since the field objects have a wide redshift range.  However, it works well as a zeroth-order correction. A more accurate result would require a full SED model for the entire redshift range of all the contaminating sources.

In addition to the \textit{GALEX} DIS, we used the deep GOODS-S survey conducted by the Ultraviolet/Optical Telescope on board the Neil Gehrels Swift Observatory \citep{Hoversten2009}.  These data, which include a 158~ksec exposure in the wide-bandpass uvw1 filter (central wavelength 2600~\AA, effective width 800~\AA), a 145~ksec exposure in wide-bandpass uvw2 (central wavelength 1930~\AA, effective width 670~\AA), and a 136~ksec exposure in the medium bandpass uvm2 filter (central wavelength 2250~\AA, effective width 530~\AA),  have a spatial resolution that is more than a factor of two higher than that of \textit{GALEX}.

To obtain the \textit{Swift} photometry, we downloaded the deep images taken of the \textit{Chandra} Deep Field South (of which GOODS-S is a region).  We then used the \href{https://github.com/UVOT-data-analysis/}{UVOT data analysis tool} and various UVOT pipeline programs to find the appropriate data in the archive, mask bad pixels, perform sky corrections, adjust individual image backgrounds, and combine the individual images into a stacked mosaic.  This mosaic was then calibrated using the \href{https://heasarc.gsfc.nasa.gov/docs/heasarc/caldb/caldb_intro.html}{HEASARC Calibration Database}.

The rest-frame wavelengths probed by the UVOT's uvw2 filter are shortward of the Lyman break (912~\AA) for all objects in our sample. We therefore used only the uvw1 and uvm2 in our analysis, performing object detections and photometry with the \texttt{UVOTDETECT} tool from the UVOT FTOOLS package (HEAsoft 6.28)\footnote{\url{https://heasarc.gsfc.nasa.gov/docs/software/lheasoft/}}.  This program is a wrapper for Source Extractor \citep{BertinArnouts1996} and was used with its default SExtractor parameters and a detection threshold of $1.2 \sigma$.  We matched the UVOT sources to the 3D-HST catalog, taking the closest measurement in each band, as long as this distance was less than $0\farcs 5$. We found 74 3D-HST galaxies with measurable NUV flux in at least one of the bands. These sources tend to have higher \Ha and \OIIIsimp fluxes and brighter rest-frame $U$ magnitudes than the typical source in our catalog.

\begin{figure}
    \centering
    \resizebox{\hsize}{!}{
    \includegraphics{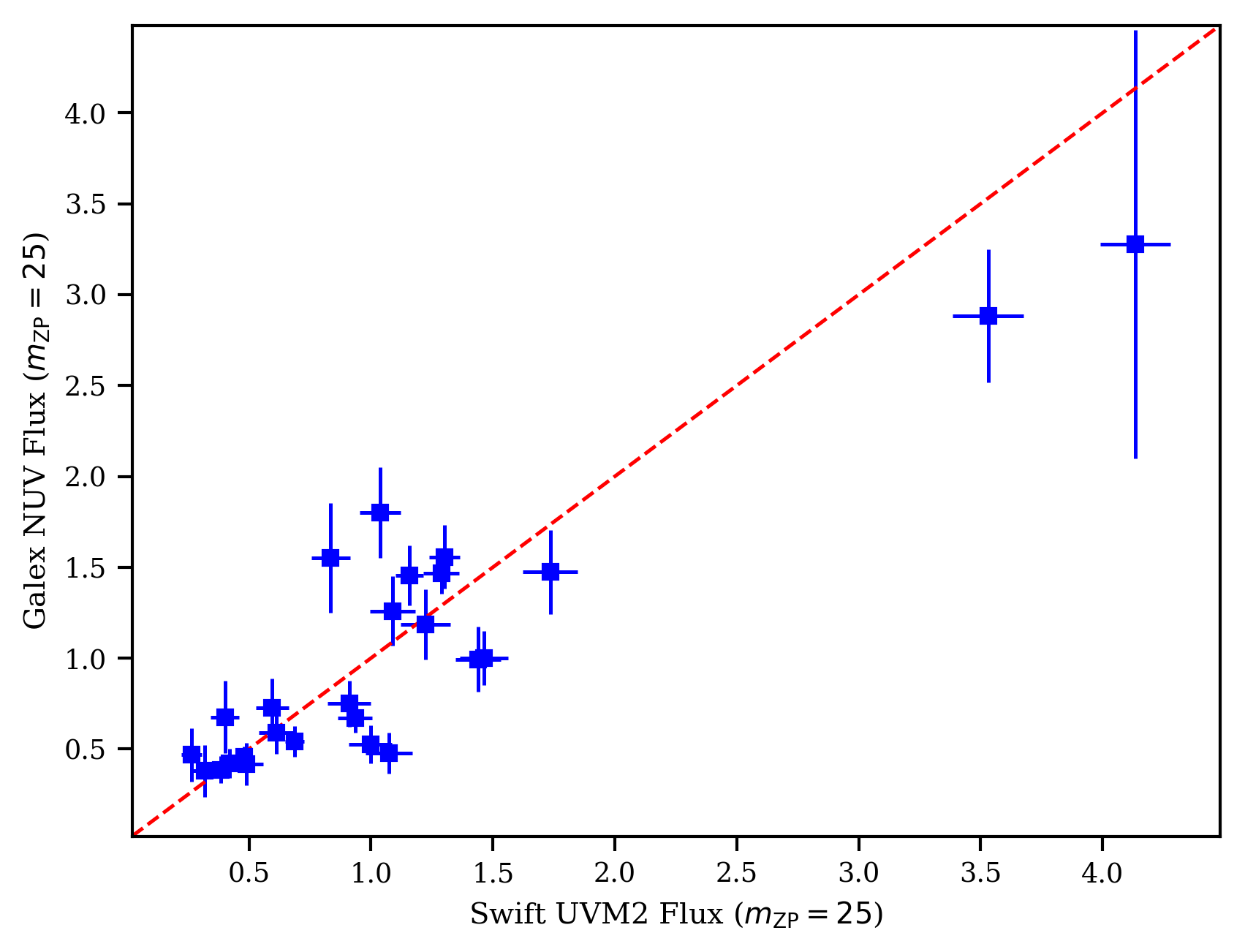}}
    \caption{\textit{GALEX} NUV ($\lambda_{\rm eff}=2304$~\AA) vs \textit{Swift} uvm2 ($\lambda_{\rm eff}=2245$~\AA) fluxes for sources with both measurements. The results are broadly consistent within error consideration.}
    \label{fig:galexvsswift}
\end{figure}

Since the central wavelength of the \textit{GALEX} NUV filter (2304~\AA) is similar to that of the textit{Swift} uvm2 filter (2245~\AA), we can use the two photometric measurements to test our approach to the problem of \textit{GALEX} NUV source confusion. The results of this comparison are displayed in Figure \ref{fig:galexvsswift}. Considering the photometric errors involved, the different bandpasses of the two instruments, and the approximation used to address the issue of \textit{GALEX} confusion, the agreement between the two measurements is very good.  This supports our use of \textit{GALEX} photometry in our SED fitting.


\subsection{\mcsed}\label{subsec:mcsed}

The most comprehensive method for transforming photometric and emission line observations into astrophysical inferences is through SED modeling \citep[see][and references therein for excellent overviews]{Walcher2011,Conroy2013}. SED modeling is effective because of the diversity of information that can be gleaned from different parts of the electromagnetic spectrum. However, the various parameters of the SED, such as the star formation history (SFH), stellar mass, dust attenuation and re-radiation, and metallicity are not completely independent \citep[e.g.,][]{Santini2015}.  As a result, galaxies with different physical conditions can have very similar SEDs, thus creating degeneracies in a complex parameter space. An effective method to efficiently and carefully examine this multidimensional space is the Markov Chain Monte Carlo (MCMC) approach (often coupled with Bayesian modeling), which plants seeds at a number of initial points in the space and grows roots (chains) from the seed locations in order to determine parameters that maximize likelihood. MCMC codes have become quite popular in astronomy, especially for the investigation of stellar populations and AGN activity in galaxies of all redshifts \citep[e.g.,][]{Acquaviva2011,Serra2011,Leja2017}.

While many SED fitting codes have been described in the literature, most are either not optimized for grism surveys with large number of galaxies, or are not designed to fit both the UV through NIR light from stars, and the mid- and far-IR emission from dust. One of the exceptions is \mcsed\footnote{\url{https://mcsed.readthedocs.io/en/latest/}} \citep{Bowman2020}, a fast, flexible SED-fitting program which uses the MCMC affine-invariant ensemble sampler \texttt{emcee} \citep{ForemanMackey2013} to efficiently investigate high dimensional parameter space.  \mcsed can fit part or all of a galaxy's SED, and can incorporate photometry, emission-line fluxes, and absorption-line spectral indices into its maximum-likelihood analysis.

In \S \ref{sec:individual}, we provide details about the \mcsed parameters used to characterize star formation histories, dust attenuation, dust emission, and metallicity.

\subsection{AGN Contamination}\label{sec:AGN}
The goal of this study is to examine the physical properties of $1.2 \lesssim z \lesssim 1.9$ star-forming galaxies, including their star-formation rates, stellar masses, dust attenuation, and (in Paper~II) dust emission.  Strong AGN activity can compromise these measurements.  Since \mcsed is not designed to model the emission from nuclear activity, the first step in our analysis was to remove obvious AGN from our sample. 

In our study, we use X-ray and NIR selection methods to identify AGN. X-ray selection is known to be quite effective given that normal galaxies tend to be limited in their X-ray emission \citep[e.g.,][and references therein]{BrandtAlexander2015}, but it fails for highly obscured sources due to X-ray absorption and scattering processes \citep[e.g.,][]{Yan2011,BrandtAlexander2015}. Conversely, NIR selection methods can identify obscured AGN that are not present in X-ray surveys, but struggle with low-luminosity and host-dominated objects \citep[e.g.,][]{Donley2012}. To truly identify all AGN lurking in our sample, we would need to combine several selection methods across a wide swath of the electromagnetic spectrum. This is beyond the scope of our study, and for the purposes of our analysis, unnecessary. Though a few of our objects may be AGN masquerading as normal galaxies, their existence will not affect the conclusions of this paper.

Deep \textit{Chandra} observations exist for all five CANDELS fields  \citep{Nandra2015,Civano2016,Luo2017, Kocevski2018,Suh2019}.  We cross-correlated our ELG sample with the list of X-ray sources, using a search radius of 1\arcsec.  Any matched object with an implied X-ray luminosity greater than $10^{42}$ \lum in the 2 to 10 keV band was considered a possible AGN and excluded from our analysis. The procedure removed 72 objects from our vetted sample of ELGs (4350 objects, see Table \ref{tab:source_counts}). 

Table~\ref{tab:AGNField} lists the number of AGN by field.  The field-to-field variation in the fraction of AGN contaminants is primarily due to the different depths of the \textit{Chandra} observations: the COSMOS data are shallowest with only 
160~ks of exposure time, while the deepest data are in GOODS-S, which has been observed for a total of 7~Ms.  We note that if adopt $10^{42}$ \lum as the threshold for AGN emission, then GOODS-S is the only field where we should be able to identify all non-Compton-thick objects. Future work with mid-IR AGN templates and more extensive IR photometry will allow for better identification of faint Compton-thick objects. 

\begin{table}[]
\hspace*{-1.4cm}\begin{tabular}{@{}lcccc@{}}
\toprule
Field  & Total & AGN & Min AGN \% & Survey Depth (ks) \\ \midrule
AEGIS  & 1196        & 15          & 1.25\%  &  800                  \\
COSMOS & 898        & 6          & 0.67\%  & 160                \\
GOODS-N & 558         & 13          & 2.33\%  & 2000                 \\
GOODS-S & 655         & 16          & 2.44\%  & 7000                 \\
UDS    & 1043        & 22          & 2.11\%  &  600                \\
All    & 4350        & 72          & 1.66\% & N/A                  \\ \bottomrule
\end{tabular}
\caption{Comparison of field-to-field AGN fraction (using the $\log L_X>42$ criterion). The values range from 0.67\% in COSMOS to 2.44\% in GOODS-S, which reflects the large range of exposure times in the different fields (last column).}
\label{tab:AGNField}
\end{table}

In addition to matching our ELGs with individual X-ray sources, we also conducted an X-ray stacking analysis on our galaxy sample.  After removing the known X-ray matches, we used the same procedure as \cite{Bowman2019} and co-added X-ray data at our sources' coordinates following the steps described by \cite{Vito2016} and \cite{Yang2017}. 

In the 2 to 10 keV band, the mean stacked luminosity for our emission line sample (excluding individually detected sources) is $6.8^{+1.2}_{-1.1} \times 10^{40}$ \lum, with the $1\sigma$ errors determined using the bootstrap method described by \cite{Yang2017}.  For comparison, the galaxies in our sample have a mean redshift of $\langle z \rangle = 1.495$, a mean stellar mass of $10^{9.5}$\msun, and a mean SFR of $12.9$ \msyr (see below).  These numbers, coupled with the conversion between X-ray luminosity and star-formation rate given by \cite{Lehmer2016}, imply an expected mean 2 to 10~keV luminosity from star formation of $\sim 4.8^{+1.0}_{-1.0} \times 10^{40}$ \lum.  In other words, there is no evidence (in X-rays) for a significant AGN population outside the known AGN\null.  Our sample therefore consists mostly of normal star-forming galaxies.

\begin{figure*}
    \centering
    \resizebox{\hsize}{!}{
    \includegraphics{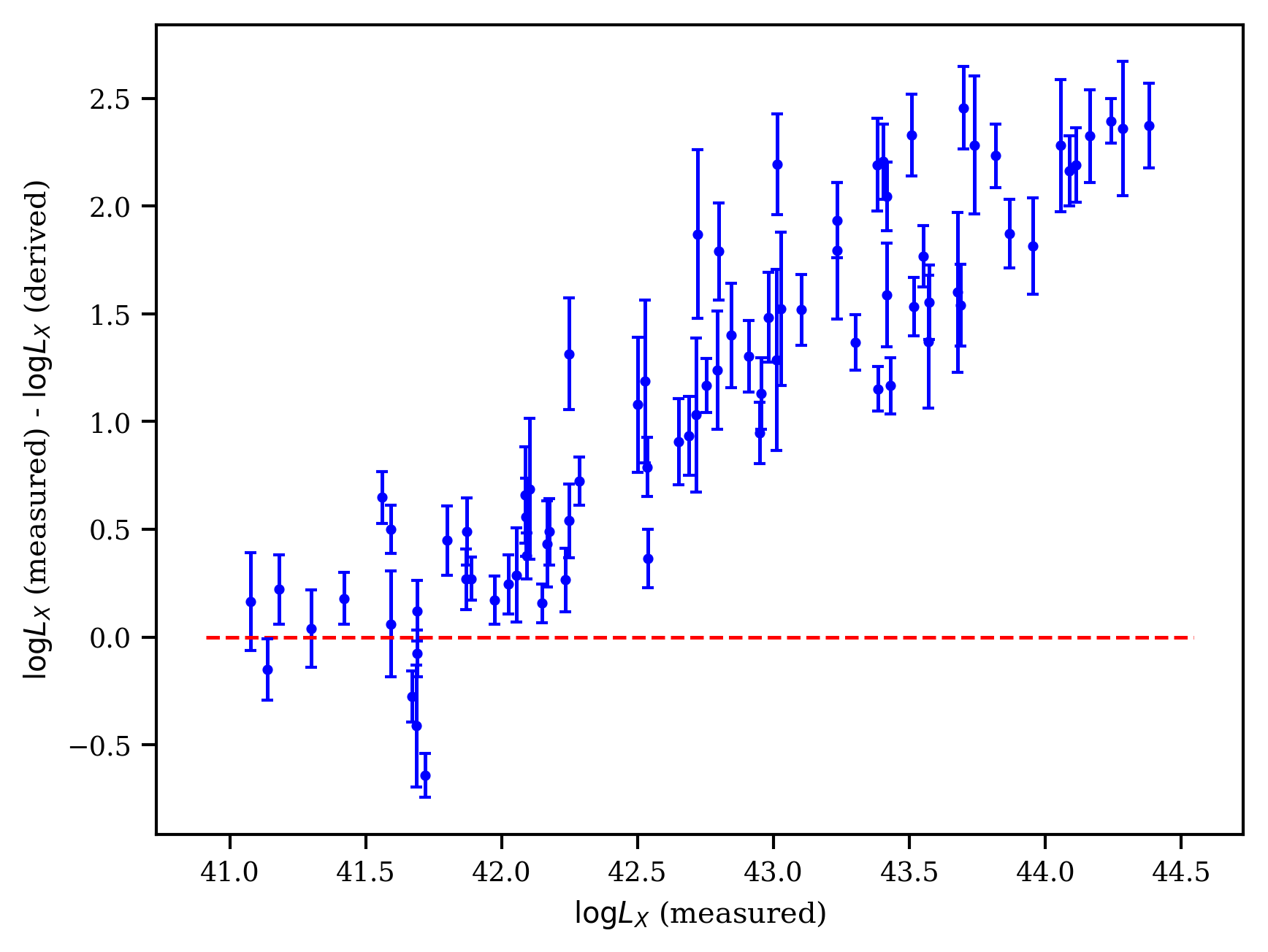}
    \includegraphics{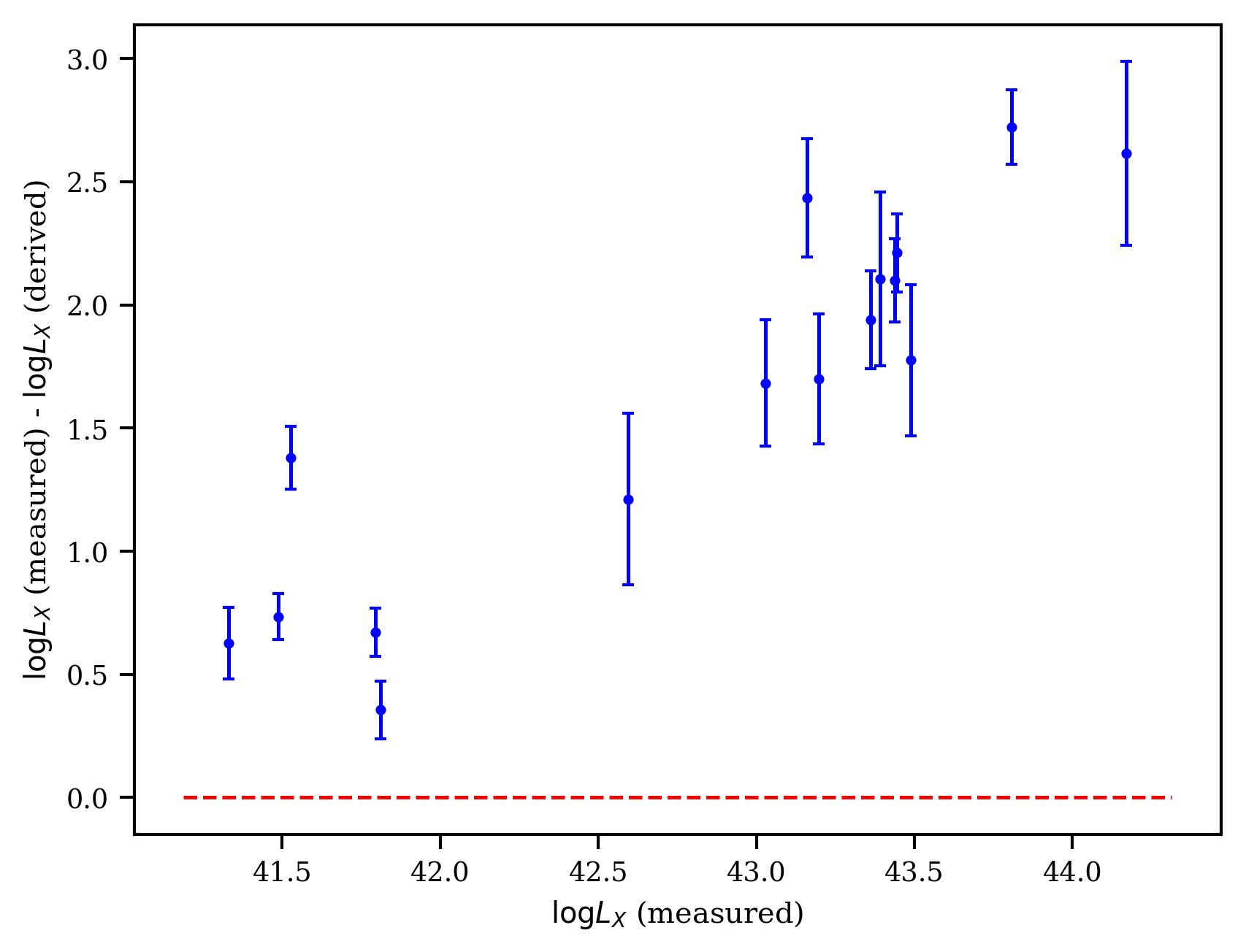}}
    \caption{Residuals for measured X-ray luminosities (2-10 keV) vs.\ derived values assuming no AGN activity. The left panel shows results for galaxies which have been detected in the mid- or far-IR; the right panel plots those objects that have no long wavelength detections.   The higher $x$-intercept of the residual curve in the left panel reflects the generally higher SFRs of the IR-bright galaxies. }
    \label{fig:AGNLx}
\end{figure*}

Figure \ref{fig:AGNLx} shows the application of the \cite{Lehmer2016} equation to all sources with X-ray measurements in our emission line sample. The left panel shows the residuals for objects detected in the mid- or far-IR by \textit{Spitzer}/MIPS or \textit{Herschel}; this is the bulk of our sample, and reflective of the fact that most AGN are also bright at mid-IR wavelengths.  The right panel shows our small sample of X-ray detected objects without a MIR/FIR measurement. The dashed line shows the relation given by \citet{Lehmer2016}. For the figures, it is clear that individual objects with X-ray luminosities above $\log L_X \gtrsim 10^{42}$ \lum have more X-ray flux than is expected from star formation.  This supports the use of our threshold for identifying AGN.

As mentioned previously, while deep X-ray surveys represent one of the most effective ways to identify AGN, very heavily obscured sources ($N_H\gtrsim (5-50) \times 10^{23}$ cm$^{-2}$) can severely reduce the observed X-ray flux to the point that they are undetected. Various methods involving IR or radio emission can be used to help identify such AGN \citep[e.g.,][and references therein]{BrandtAlexander2015}. Using the \cite{Donley2012} IRAC-band test for AGN selection, we found 39 AGN in our sample that were not identified by the X-ray surveys. Eleven sources were identified by both methods. We find that including or not including the Donley AGN in our sample makes no difference in the empirical relations between stellar mass and absolute JH magnitude presented in \S \ref{sec:massabsmag}.

\subsection{Comparison to Photometric Redshift Sample}\label{sec:emvsphot}


In \S \ref{sec:massabsmag} we introduce a powerful empirical relation between stellar mass and JH magnitude. In order to test the universality of the relation, we need a different set of galaxies, covering the same redshift range, but with a different selection criteria.  To create this catalog,  we identified a sample of 19,289 3D-HST galaxies whose redshifts were determined solely by their photometry. (These objects have either no noticeable emission lines or have contaminated/poor grism spectra.) Following the steps in \cite{Bowman2019}, we identified galaxies with photometric redshifts $1.16<z_{\rm{phot}}<1.90$ marked as reliable in the \cite{Skelton2014} catalog and with 68\% redshift confidence intervals $|z_{\rm{84}}-z_{\rm{16}}|<0.3$. 

In order to characterize the accuracy of the photometric redshifts, we compared the photo-$z$'s to ground-based spectroscopic measurements for 362 objects measured in the same manner as for the ELG sample (see \S \ref{subsec:sample}). Based on Equation \ref{eq:sigmanmad} the average error and the outlier definition are $\sigma_{\rm{NMAD}}=0.026$ and 20.4\% respectively. For comparison, we reiterate the corresponding values for the ELG sample are $\sigma_{\rm{NMAD}}=0.0017$ and 1.4\%. In other words, the photometric sample has over an order of magnitude more uncertainty in the redshift estimation.

The photo-$z$ sources with spectroscopic redshifts span a smaller range of physical properties than the general sample.  The most obvious difference is in their apparent magnitudes:  because most photo-$z$ galaxies are not ELGs, the their spectroscopic redshifts rely more on absorption features than emission lines.  Thus, while the median JH magnitude of the full photo-$z$ sample is 24.7, that for the spectroscopic sub-sample is 22.5. In this bright range, there is no correlation between redshift error and apparent magnitude, which suggests that the average error and outlier fraction quoted above may apply to the full sample.  In any case, the precise values for these numbers do not have a strong effect on our analysis.

\section{Galaxy SED Fitting} \label{sec:individual}
We inferred the physical properties of the CANDELS fields ELGs by fitting the galaxies' rest-frame UV through NIR SEDs with \mcsed using the following assumptions:

$\bullet$  The stellar emission is modeled using a linear combination of Simple Stellar Population (SSP) SEDs taken from the Flexible Stellar Population Synthesis (FSPS) library \citep{Conroy2009, Conroy2010SPSM} using Padova isochrones \citep{Bertelli1994,Girardi2000, Marigo2008} and a \citet{Kroupa01} initial mass function.

$\bullet$  Emission from ionized gas is computed using a grid of \cloudy models \citep{Ferland1998,Ferland2013} computed by \citet{Byler2017} from the above-mentioned Padova isochrones.  This grid lists both nebular continuum and line-emission as a function of metallicity, ionization parameter, and age.

$\bullet$ A star-formation rate history is defined using 6 age bins. In terms of log years, the edges of the bins are $[8,8.5,9,9.5,9.8,10.12]$. Within each bin, the SFR is assumed to be constant.

$\bullet$ The attenuation law follows the prescription of \cite{Noll2009} and \cite{Kriek2013}.  This ``Noll law,'' which generalizes the equations in \citet{Calzetti2000}, has three parameters: the total amount of stellar attenuation, $E(B-V)$, the strength of the excess attenuation bump at 2175~\AA, and the difference between the wavelength dependence in the UV and that given by the \citet{Calzetti2000} law, i.e., the ``UV slope'', $\delta$. \mcsed assumes that the dust attenuation around older stellar populations (stellar) is a constant times the attenuation around birth clouds (nebular). We used the value of $0.44$ from \cite{Calzetti2000}, but the value is easily adjustable.

\mcsed can accept a variety of input data, including broad-, medium-, and narrow-band photometry, emission-line strengths, and absorption line spectral indices.  For the case of our $1.2 < z < 1.9$ ELGs, emission-line fluxes can provide powerful constraints on a galaxy's star-formation rate and present-day metallicity.  However, not all emission-lines have the same probative value:  while recombination lines such as \Ha and \Hb are directly tied to the flux of ionizing photons and therefore a galaxy's star-formation rate, the strengths of collisionally-excited lines, such as \OIII, depend on a number of other parameters, such as the metallicity and ionization parameter of the interstellar medium (ISM). Thus, \mcsed allows the user to specify the relative weights of emission-line measurements for the $\chi^2$ calculation.

In our computations, we assigned the \OIII emission line to have the same weight as a photometric measurement. Even though \OIIIsimp tends to be very strong in the sample, the physics behind its strength is complicated and the line may not be accurately reproduced in our grid of \cloudy models \citep[see the discussion in][]{Bowman2020}. On the other hand, we gave \Ha, which is nearly ubiquitous in sources with $z < 1.55$, 2.5 times the weight of a photometric measurement.  

Meanwhile, while the physics of \Hb is equally well understood, this line tends to be weak and is more affected by underlying stellar absorption than \Ha. On the other hand, a large fraction of $1.2<z<1.5$ galaxies in the \cite{Momcheva2016} catalog have unphysical Balmer decrements, suggesting that either \Hb is being overestimated (perhaps through an underestimation of the continuum level or over-correction for absorption) or \Ha is being underestimated, with the former more likely. In any case, \Hb tends to lie under the sensitivity limit (Figure \ref{fig:HaOIIIcomplete}), so we gave \Hb a weight of just one photometric measurement.  

The precise weights given to the emission-line measurements do not substantially affect our results. For example, changing the \Ha weight from 2.5 to 5 or the \Hb weight from 1 to 2 leads to SED-derived parameter distributions (SFR, stellar mass, etc.) that are statistically indistinguishable (through K-S tests) from our default weighting.

Motivated by the strong \OIII fluxes (e.g., right panel of Figure \ref{fig:HaHb}), we assigned a high ionization parameter to the galaxies in our sample: $\log U=-2.5$. \cite{Bowman2020} discuss this choice in detail. From their Figure 14, we observe that regardless of the gas-phase metallicity, $\log U$ must be $\gtrsim -2.5$ if we are to reproduce the \OIIIsimp/\Hb ratios observed in the ELGs between $1.2 < z < 1.9$.  Setting $\log U=-2.0$ does not have a large impact on our results, and values any higher than this would not be reasonable for non-AGN. 

\begin{figure*}
    \centering
    \includegraphics[width=1.0\textwidth]{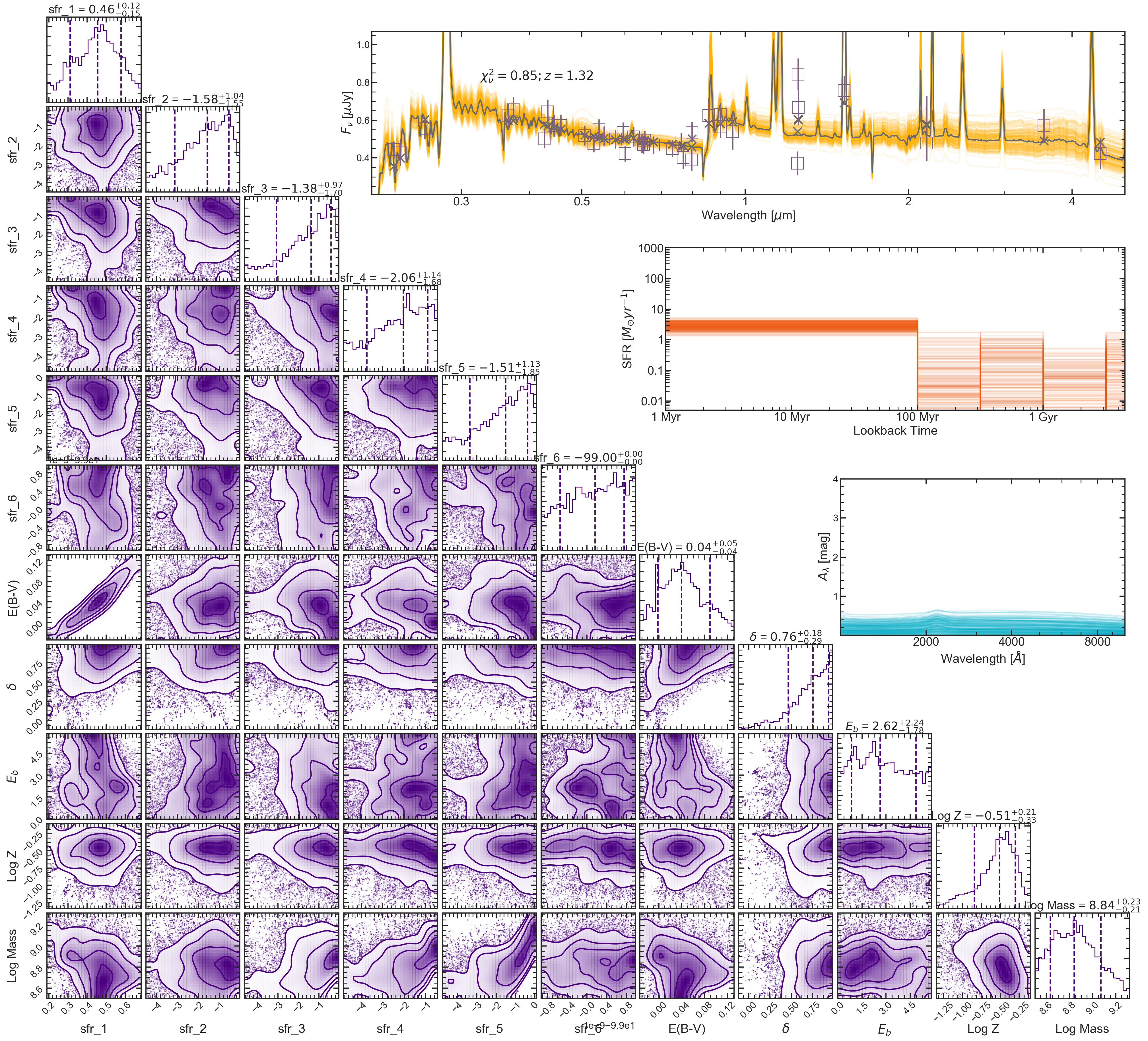}
    \caption{A graphic summary of the results of SED fitting for the source GOODS-S 3178.  This fit uses a 6 bin SFH, the Noll dust attenuation law, and  metallicity as a free parameter. The corner (triangular) plots show every 2D histogram of the explored MCMC parameter space. At the top of each column is the distribution of values for the given parameter marginalized over the entire MCMC process.  At the top right is the observed SED (open black boxes) with 200 (out of 100,000) randomly chosen realizations (yellow) of the derived SED from the MCMC parameters. The median SED is shown in black, and the modeled flux density values are shown with "x" marks. Below that panel is the SFR as a function of look-back time for the 200 aforementioned random realizations of parameter values. Finally, the third plot on the right shows the 200 realizations of the dust attenuation curve (as a function of wavelength).}
    \label{fig:trianglever2}
\end{figure*}

We then used \mcsed to fit the SEDs of our galaxies, using 100 walkers (random initial points in parameter space) and 1,000 chains (explorations of the parameter space from each initial point) to map out the phase-space likelihoods. 
As an example of our \mcsed fits, we show in Figure~\ref{fig:trianglever2} the ``triangle plots'' for source GOODS-S 3178 ($z=1.24$) using the assumptions laid out above. 

Included in the triangle plot are the various two-dimensional projections of the likelihood space, the marginalized likelihood distributions for each parameter, the best-fit SED (along with a set of 200 realizations that are randomly drawn from the parameter posterior distributions), the star-formation rates in our age bins, and the derived dust attenuation curve.  

The figure demonstrates the complicated nature of galaxy SED modeling. Some parameters, such as metallicity, are well-defined (primarily from the emission-line constraints); some, such as the 2175~\AA\ bump strength, are essentially unconstrained; and some, such as amount of internal reddening, show a strong degeneracy with other parameters (in this case, the recent star-formation rate). The figure suggests that although the SFR of past epochs is poorly constrained, for most of its history, the galaxy has gradually increased its rate of star formation until the epoch of observation, when $\mathrm{SFR}\sim 2.6~\msun~\mathrm{yr}^{-1}$.  The figure also shows that there is little to no internal stellar attenuation in this galaxy, with $E(B-V) = 0.03^{+0.06}_{-0.04}$, and there is a suggestion of a 2175~\AA\ bump, though the distribution of values is quite broad.

\subsection{Mass-Metallicity Relationship} \label{sec:emprelmass}

While \mcsed provides substantial insight into individual sources, the general trends we find for all of galaxies in the sample are much more scientifically interesting. The first of these is the mass-metallicity relationship that comes from the best-fit \mcsed parameters.

The metallicities derived from \mcsed are technically stellar metallicities, based on the collection of SSPs gathered to create a complex stellar population whose SED best fits the data. However, considering most of our sources are star forming galaxies, the photometry that inform the metallicity fit are largely influenced by light from young stars, whose metallicities are nearly identical to that of the ISM given the short timescales involved.

In addition, the SSP grid includes nebular emission, inferred from an interpolation over \texttt{CLOUDY} tables with a fixed ionization parameter \citep{Byler2017}. This is heavily influenced by the emission line strengths input into \mcsed. In other words, even though \mcsed reports stellar metallicities, for star forming galaxies, these are likely pretty similar to the gas-phase metallicities. For this reason, in \mcsed we set the gas-phase metallicity equal to the stellar metallicity.

\begin{figure}
    \centering
    \resizebox{\hsize}{!}{
    \includegraphics{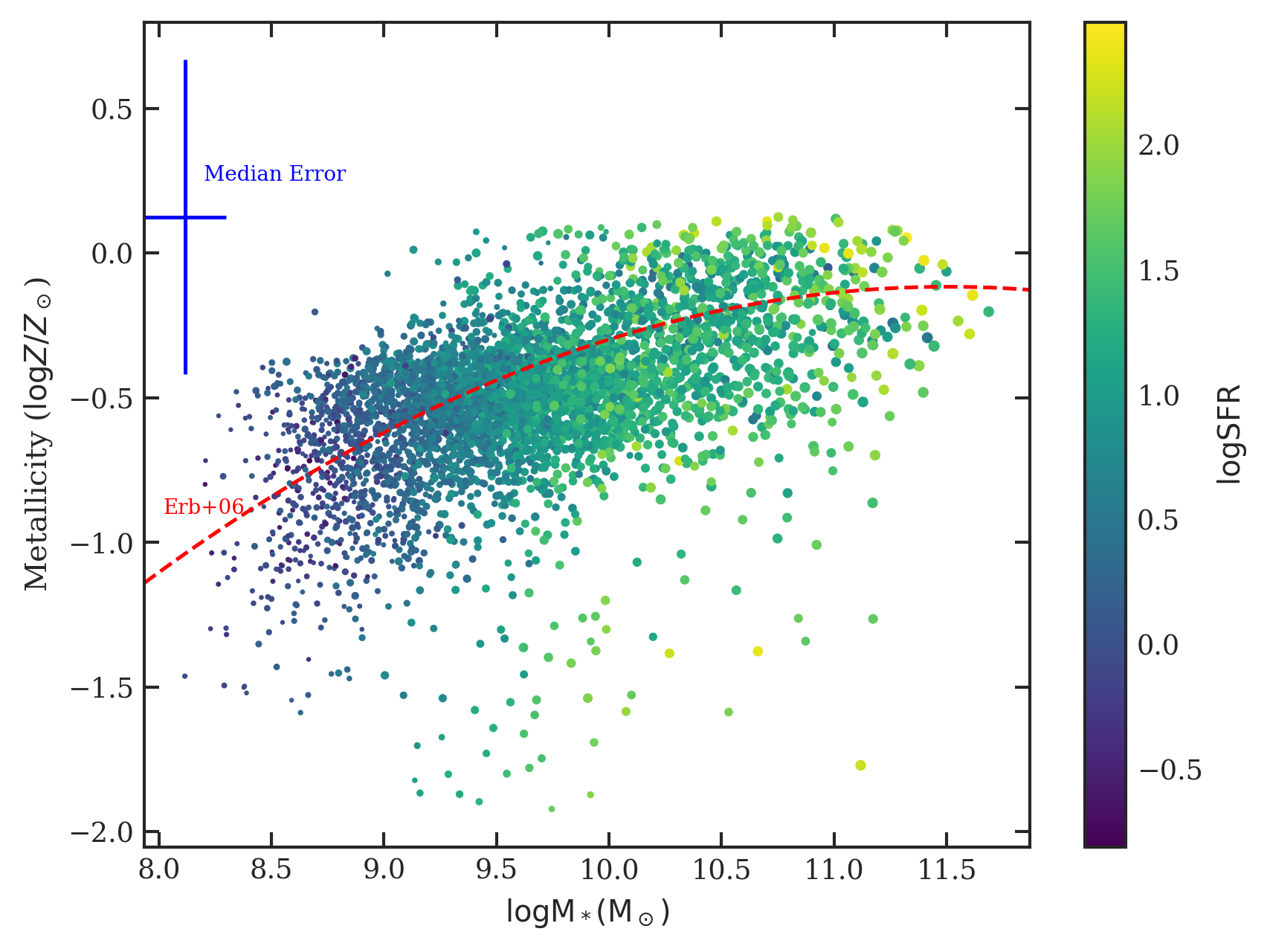}}
    \caption{Metallicity vs stellar mass. While the scatter is large, the trend with stellar mass are clear. The agreement with the mass-metallicity relationship from \cite{Erb2006} is quite striking. See text for discussion about stellar vs gas-phase metallicity. The points are sized based on JH magnitude. The points are colored by $\log$ SFR.}
    \label{fig:massmetsize}
\end{figure}

We show metallicity vs.\  stellar mass in Figure \ref{fig:massmetsize}. It is clear that our data conforms very well to the high redshift mass-(gas-phase) metallicity relationship from \cite{Erb2006} as modified from \cite{Tremonti2004}, although the scatter is quite large. 


\section{Stellar Mass and NIR Absolute Magnitude} \label{sec:massabsmag}



As discussed in \S \ref{sec:emprelmass}, the most interesting results we can gain through SED fitting are empirical trends in physical and observational quantities. The mass-metallicity relation is useful in the sense that it provides physical insight into how galaxies evolve: galaxies' ISM tends to get enriched as they grow in mass. The details of that enrichment depend on the complex interplay between stars, gas, dust, and the galaxies' external environment. The ability to reproduce scaling relations is important for models that are built from first principles.

Another type of useful relation is one that allows for computational shortcuts.  As discussed in \S \ref{sec:intro}, an empirical relation that can rapidly measure stellar masses for millions of galaxies will be very useful for future missions like \textit{Euclid} and \textit{Roman}.

The parameter that best traces the stellar mass is the absolute JH magnitude. To obtain this quantity, however, we had to apply a K-correction to the apparent magnitude; this factor was computed using each galaxy's best-fit \mcsed spectrum and the response functions for the F125W, F140W, and F160W filters. For completeness, we provide the equations used for the K-correction below \citep{OkeSandage1968}. 


\begin{align} \label{eq:kcorr}
    &m_{JH} = M_{JH} + 5\log \left(\frac{D_L}{10~{\rm{pc}}} \right) + K + A_{JH} \\
    &K = -2.5\log (1+z) - 2.5\log \left(\frac{\int F_\nu[\nu (1+z)]S_{JH}(\nu)d\nu}{\int F_\nu(\nu)S_{JH}(\nu)d\nu} \right)
\end{align}
Here $D_L$ is the luminosity distance, $S_{JH}$ is the response curve of the (J+JH+H) bandpass, and $A_{JH}$ is the average foreground extinction in the bandpass, which we assume to be zero given the long wavelengths involved and the high galactic latitudes of the CANDELS fields.  When this factor is applied, the stellar mass becomes related to the flux in the JH band (which combines the F125W, F140W, and F160W filters) by a simple power law.
Specifically, 

\begin{equation}
    \log M_* = (-0.393 \pm 0.002) M_{JH} + 1.22 \pm 0.04
\label{eq:jhabs}
\end{equation}

or 

\begin{equation}
    M_* \propto F_{{\rm JH,abs}}^{0.98}
\end{equation}

This relation accounts for 93\% of the variance between the two variables ($R^2=0.93$) and has a low residual standard error of $0.17$ dex. The relation is good for all sources with absolute magnitudes fainter than $M_{JH}=-26$. 

In fact, very similar relations exist for each individual infrared band, as the fits for all three bandpasses have comparable values of $R^2$ ($0.91-0.93$) and residual standard errors between 0.16 and 0.19~dex. The slightly larger error for the J band is likely due to the effect of ongoing star formation, which is stronger at shorter rest-frame wavelengths.
\begin{align} \label{eq:absmags}
    \log M_* &= (-0.410 \pm 0.002) M_{{\rm{F125W}}} + 0.85 \pm 0.04 \nonumber \\
    \log M_* &= (-0.399 \pm 0.002) M_{{\rm{F140W}}} + 1.05 \pm 0.04 \\
    \log M_* &= (-0.390 \pm 0.002) M_{{\rm{F160W}}} + 1.22 \pm 0.04 \nonumber
\end{align}
Note that as the observed wavelength range in Equation \ref{eq:absmags} gets redder, the magnitude coefficient gets smaller (less negative), meaning that the stellar mass is proportional to a slowly decreasing power of the band luminosity, with the exponent around $1$ (i.e., direct proportionality).

\begin{figure*}
    \centering
    \resizebox{\hsize}{!}{
    \includegraphics{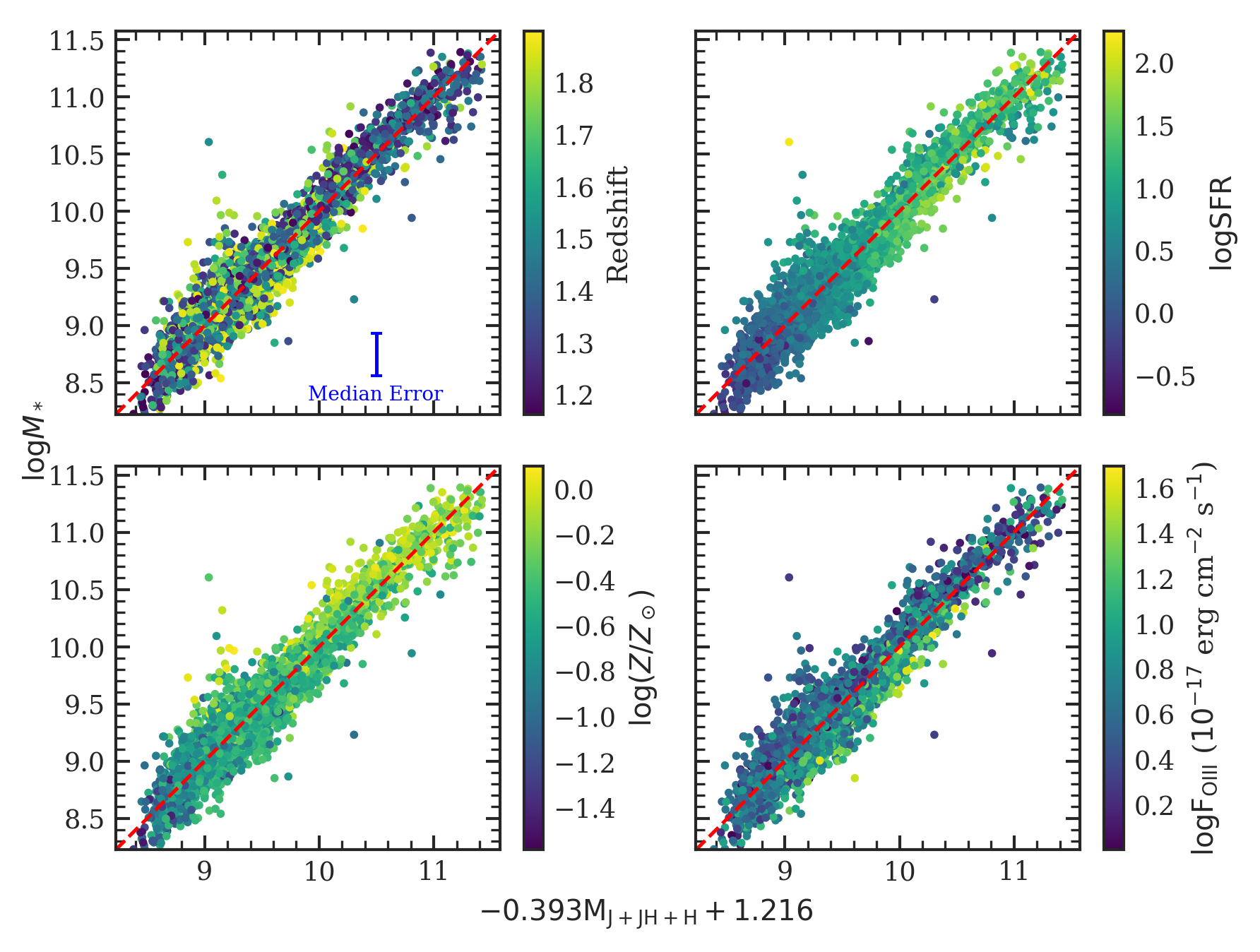}}
    \caption{The relation between stellar mass and absolute JH magnitude as written in Equation \ref{eq:jhabs}. The correlation is quite tight (scatter of $\sim 0.17$ dex). Each panel shows the points colored by a different quantity: redshift (top left), SFR (top right), \OIIIsimp flux (bottom right), and metallicity (bottom left). Stellar mass has virtually no dependence on the other these four parameters.}
    \label{fig:massvsjhabs}
\end{figure*}

In Figure \ref{fig:massvsjhabs}, we show the how stellar mass varies with absolute JH magnitude, with points colored by redshift, sSFR, \OIII flux, and metallicity. From the figure, it is clear that the absolute flux contained in this one wide bandpass (F125W + F140W + F160W) --- or in any one of the three NIR bandpasses --- does a remarkable job of predicting an ELG's stellar mass: the average scatter about the best-fit line is only $\sim 0.17$ dex.  Moreover, no single parameter dominates the dispersion of the fit; the four parameters tested all scatter equally about the best-fit line. 

The lack of correlation between the properties coloring the points and the direction perpendicular to the relation suggests that neither redshift, metallicity, SFR, or \OIII flux contribute significantly to the scatter.  The only parameter that may have a small systematic is metallicity: low metallicity galaxies tend to be slightly brighter in the rest-frame optical/NIR at the same stellar mass.  This is likely due to the effects of opacity, as with few metals, the flux distribution of main-sequence and red giants are shifted slightly to bluer wavelengths and the stars are slightly brighter.

\begin{figure}
    \centering
    \resizebox{\hsize}{!}{
    \includegraphics{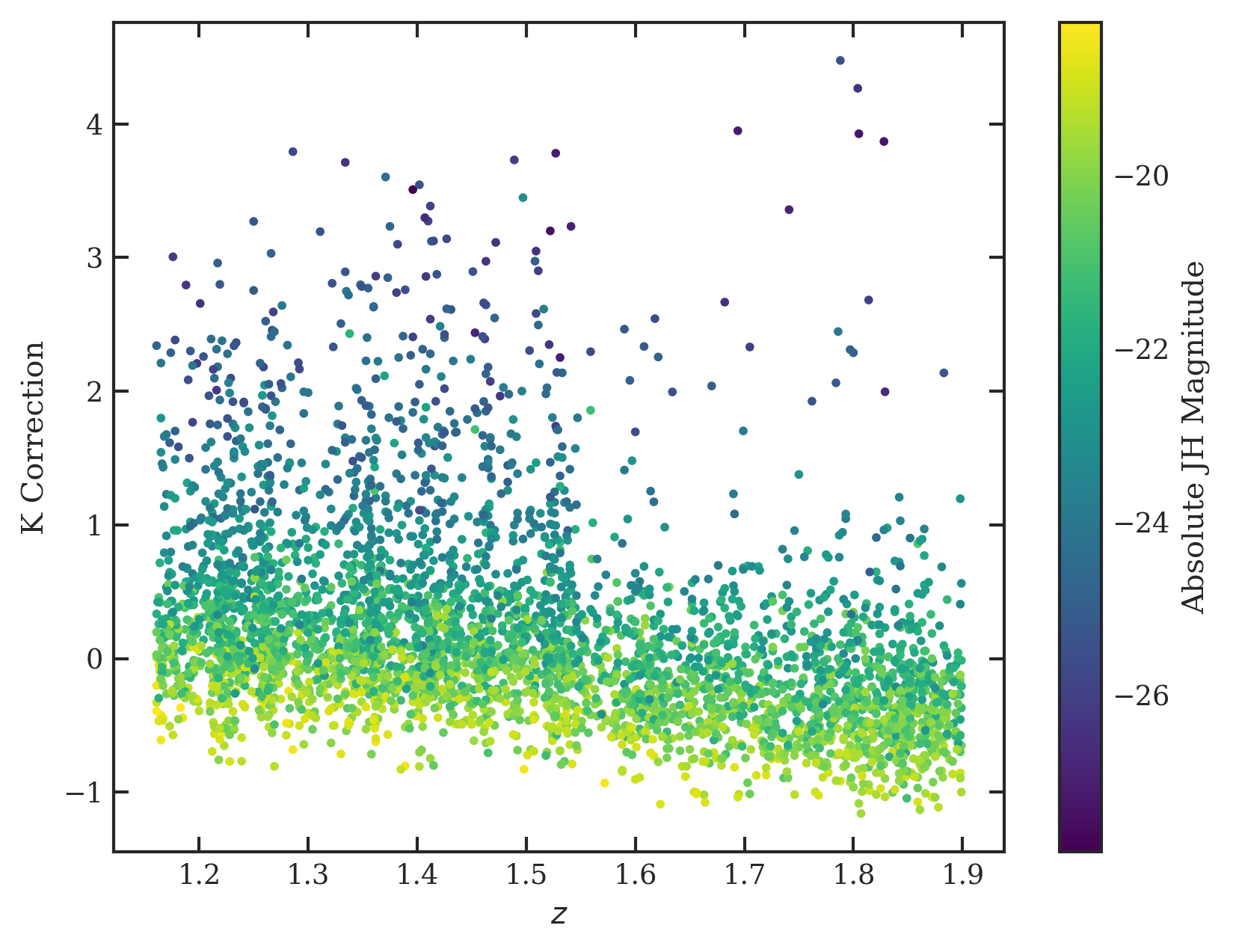}}
    \caption{K-correction as a function of redshift. The points are colored by absolute JH magnitude. There is a small but clear trend with redshift. Note that the scatter in the diagram is mostly due to the absolute JH magnitude itself: at a given redshift, brighter sources tend to have larger K-corrections.}
    \label{fig:Kcorr}
\end{figure}

While our rather large redshift range ($1.2 < z < 1.9$) necessitated the application of a K-correction, at NIR wavelengths this factor is rather insensitive to redshift (see  Figure \ref{fig:Kcorr}). Therefore, Equation \ref{eq:jhabs} is highly useful for rough stellar mass calculations. Moreover, it is quite simple to interpret. Even in star forming galaxies, the vast majority of stars are not UV-bright but rather emit most of their light in the red and NIR\null. In other words, while young, massive stars dominate the light in the rest-frame ultraviolet, their effect on a galaxy's red and NIR emission is minor. Thus, at these wavelengths, the bulk of the emission comes from the much larger population of older, lower-mass stars.  By measuring this light, we are probing the stellar mass. Again, what is surprising is how tight the relationship is over such a large redshift range and diverse set of galaxies.

\begin{figure*}
    \centering
    \resizebox{\hsize}{!}{
    \includegraphics{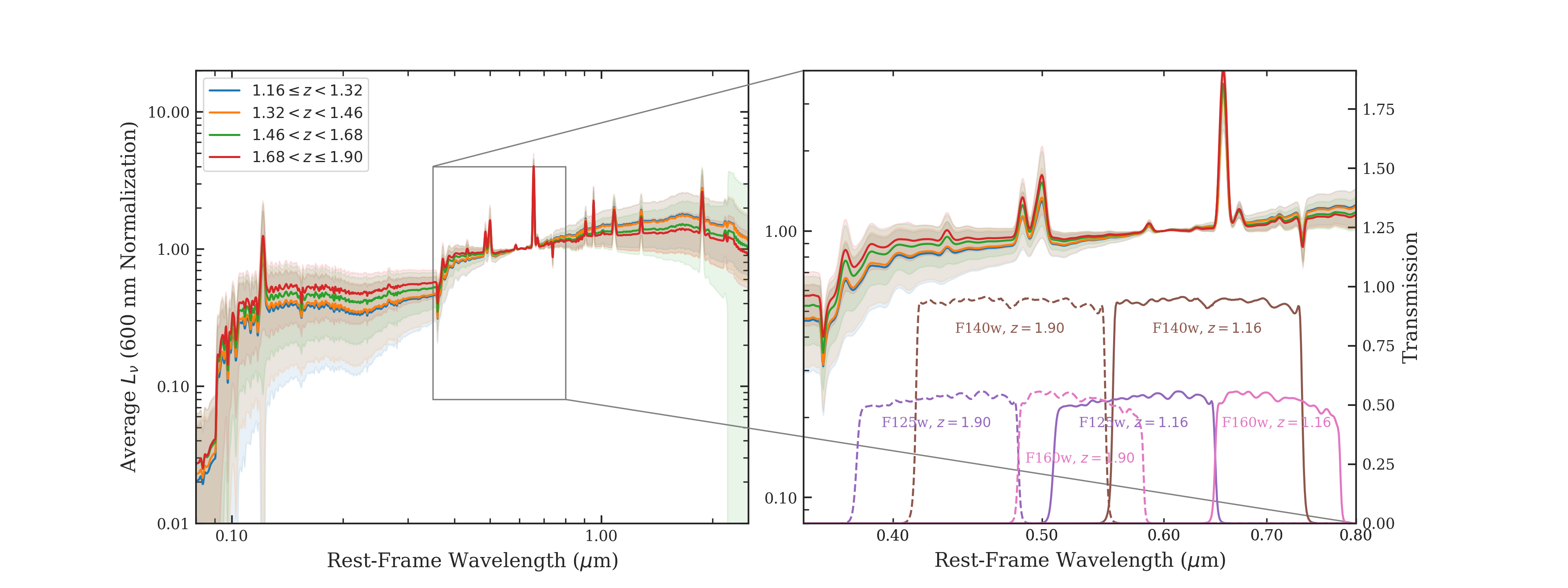}}
    \caption{\mcsed-derived energy distributions for our ELG sample. The data are normalized at 6000~\AA\ and then stacked into four redshift bins. There is very little SED evolution over our redshift range. In addition, as shown in the inset (right), the specific luminosity is slowly increasing with wavelength in the rest-frame optical, a trend that continues into the NIR (as seen in the left panel); this tends to cancel the $(1+z)$ effect in the K-correction. Together, these two observations help explain why the stellar mass-absolute magnitude relation is so tight.}
    \label{fig:stackspec}
\end{figure*}

Since we are using a single NIR band to observe galaxies over a range of redshifts, the rest-frame bands (before the application of a K-correction) are changing with redshift: the higher the redshift, the bluer the band. But, as Figure \ref{fig:massvsjhabs} illustrates, Equation \ref{eq:jhabs} depends only weakly on redshift. To explain this phenomenon, we stacked the \mcsed best-fit spectra as a function of redshift (Figure \ref{fig:stackspec}). Based on these spectra and the K-corrections in Figure \ref{fig:Kcorr}, we identify two effects that conspire to produce the tight correlation: 1) the almost negligible SED evolution in the rest-frame visible and NIR over the redshift range in question and 2) the fact that most ELGs are strongly star forming, and hence have a slightly positive optical spectral slope ($F_\nu$ vs.~$\lambda$) that is nearly independent of redshift. This slope almost cancels the $(1+z)$ term in the equation for the K-correction. 

In contrast, quiescent galaxies have steep positive optical slopes which overcome the $(1+z)$ term; this leads to a large positive K correction and a strong redshift dependence. For this reason, formulating a relation like Equation \ref{eq:jhabs} for quiescent galaxies using an observed-frame absolute magnitude is ill-advised.

Note that if we increase the redshift to $z\sim 2.5$, the J+JH+H band will be centered around the 4000 \AA\ break, where the spectral slope rapidly changes.  At this point, it is likely that Equation \ref{eq:jhabs} breaks down. Later in this section, we discuss the analog to this relation at lower redshifts.


We demonstrate the balance between the small positive NIR spectral slope and the $(1+z)$ term in Figure \ref{fig:Kcorr}, where we plot evolution of K-correction over the redshift range of our survey.  For the most part the K-corrections are small. While a trend with wavelength is clearly visible, it is minor, and if we take this slope into account, almost all the residual redshift dependence of Equation \ref{eq:jhabs} is gone. As a result, the relations shown in Figure \ref{fig:massvsjhabs} are virtually independent of redshift.  

In Figure \ref{fig:Kcorr}, we notice that at a given redshift, brighter sources generally have larger K-corrections. This arises from three effects. First, higher-mass galaxies tend to have more dust \citep[e.g.,][]{Pannella2015,Bogdanoska2020}, which serves to shift the energy balance to redder wavelengths. Second, the mass-metallicity relation \citep[e.g.,][]{Erb2006} implies that brighter, higher-mass galaxies have more metal-rich, redder populations. Finally, and perhaps most importantly, higher-mass galaxies are often quenched or are about to be quenched.  Thus, these objects have smaller sSFRs than the lower-mass systems.

\begin{figure*}
    \centering
    \resizebox{\hsize}{!}{
    \includegraphics{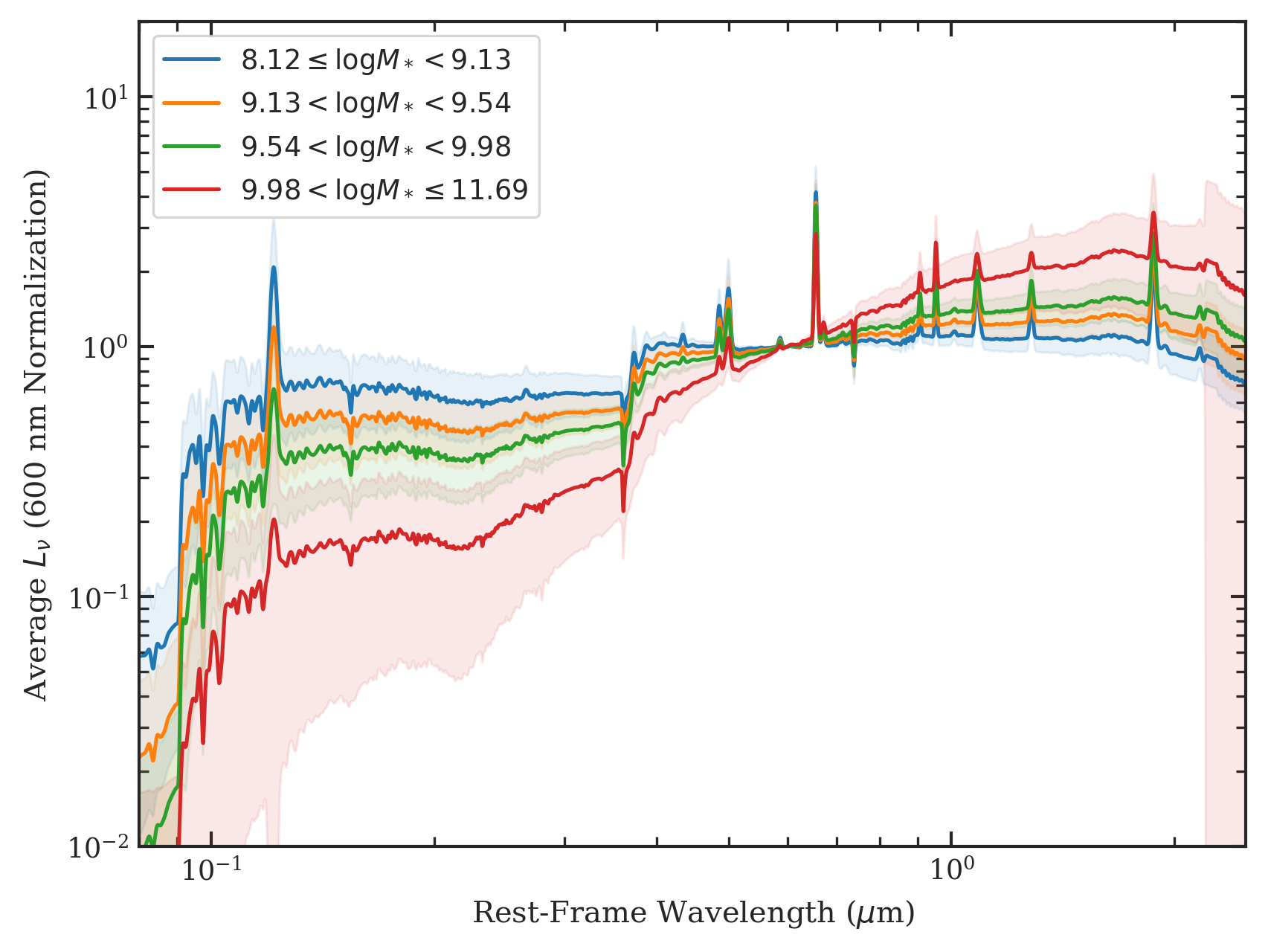}
    \includegraphics{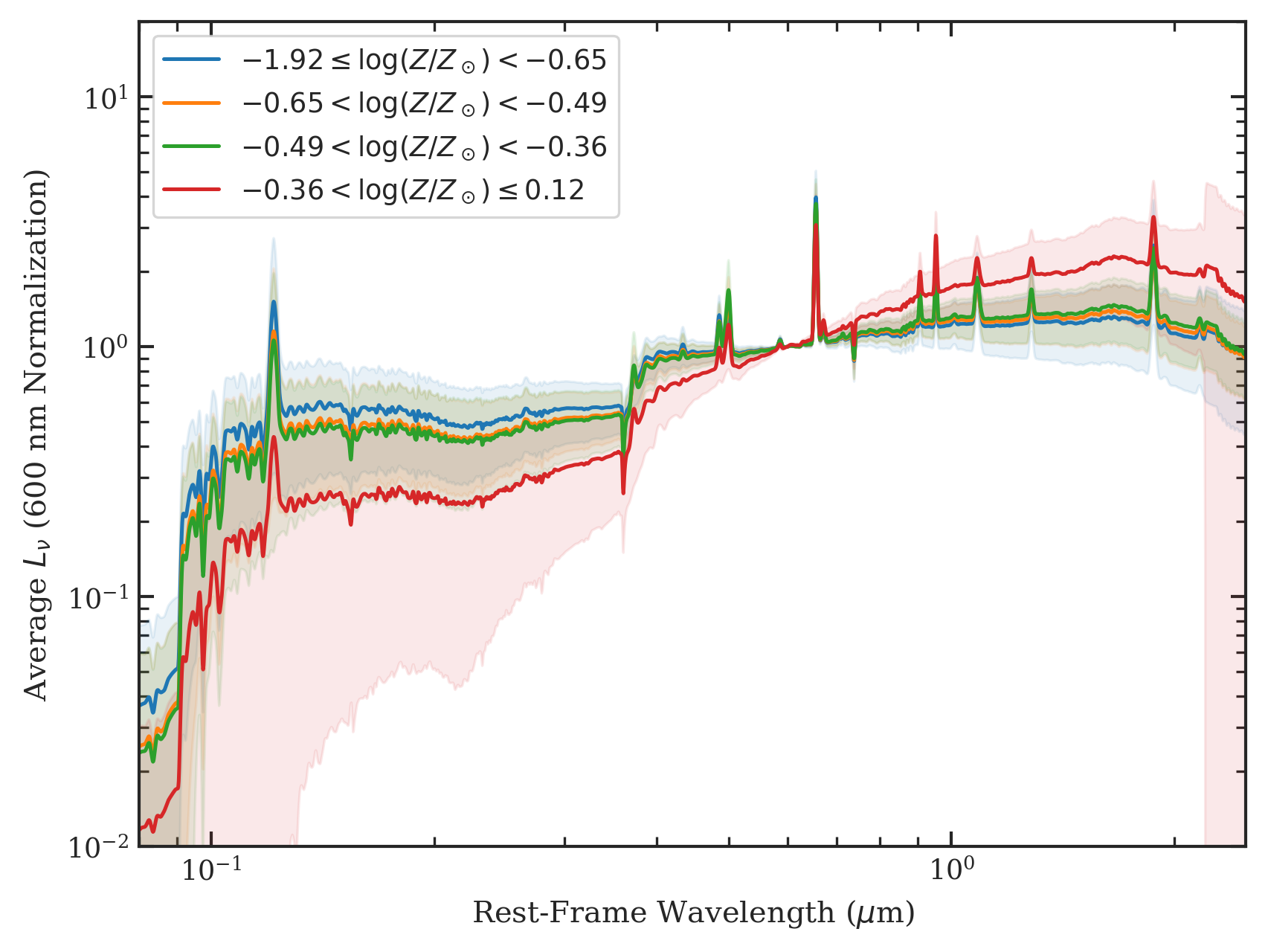}}
    \resizebox{\hsize}{!}{
    \includegraphics{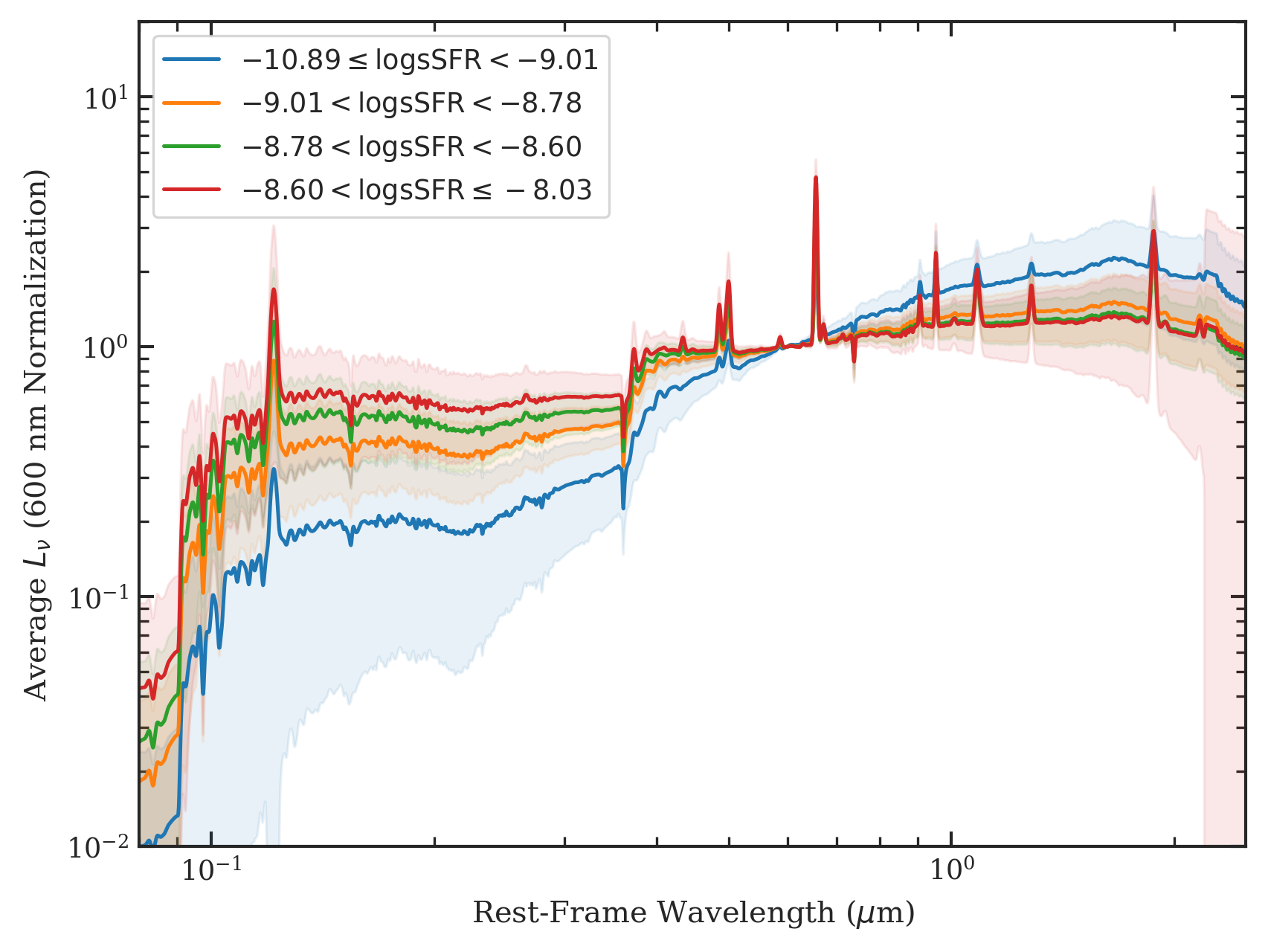}
    \includegraphics{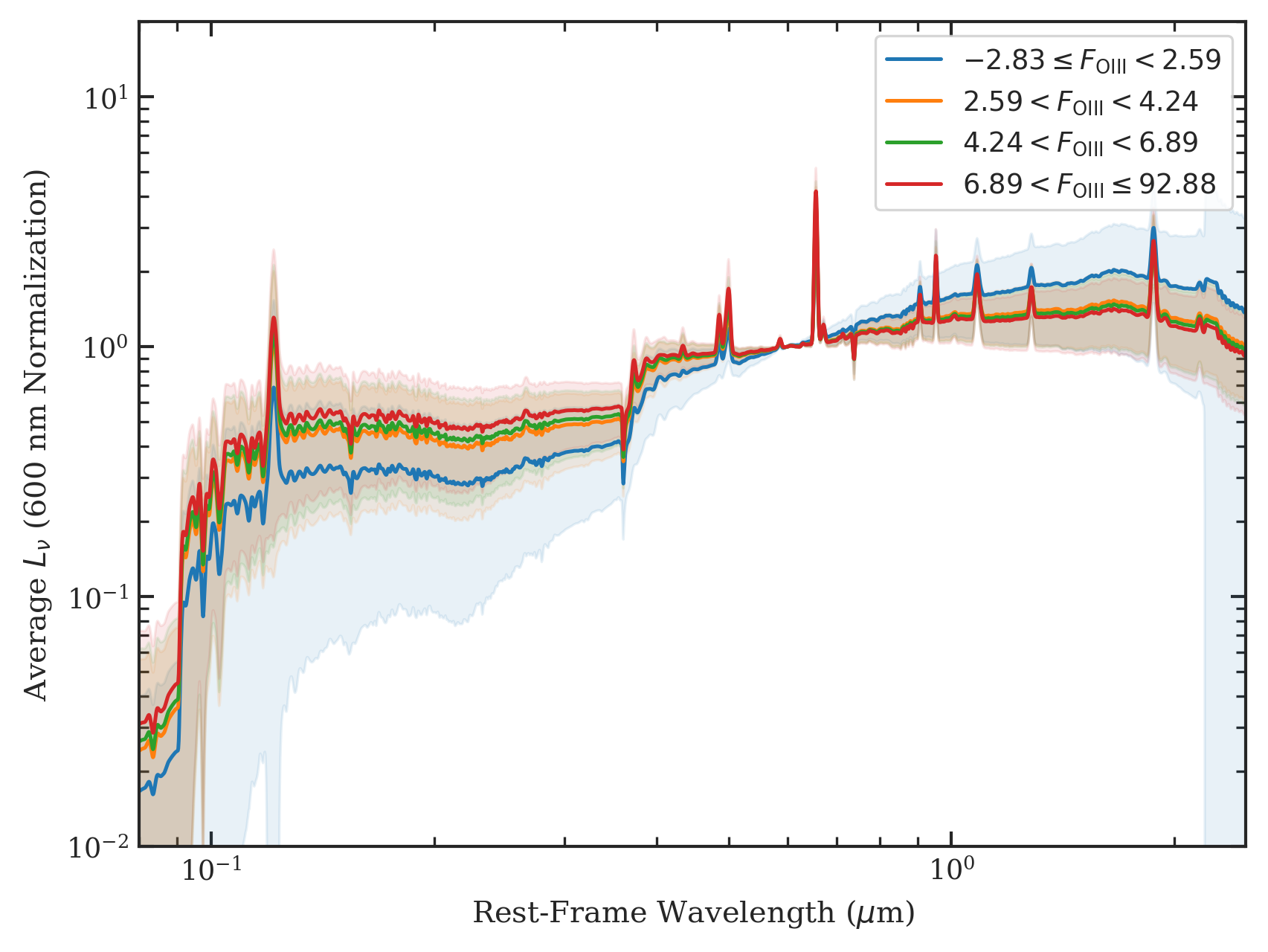}}
    \caption{\mcsed-derived spectra for our ELG sample normalized by the 6000~\AA\ flux and stacked by stellar mass (top left), metallicity (top right), sSFR (bottom left), and \OIIIsimp flux (bottom right). It is clear that the most massive galaxies, which tend to also have the highest metallicities, lowest sSFRs, and lowest \OIIIsimp fluxes, have redder spectra than their lower-mass counterparts. }
    \label{fig:stackspecextra}
\end{figure*}

These effects are illustrated by Figure \ref{fig:stackspecextra}, in which we stack the best-fit \mcsed spectra by stellar mass (top left), metallicity (top right), \OIII flux (bottom right), and sSFR (bottom left). Based on the similarities of each set of four SED shapes, we conclude that the highest mass galaxies generally have the highest metallicities, lowest \OIII fluxes, lowest sSFRs, and reddest spectra.

As we discussed earlier, the redder spectra associated with massive and quiescent galaxies broadens the relation between stellar mass and absolute magnitude.  Therefore, we restricted the range of absolute magnitudes considered in the mass-magnitude relations to galaxies fainter than $M_{J+JH+H}>-26$.


\begin{figure}
    \centering
    \resizebox{\hsize}{!}{
    \includegraphics{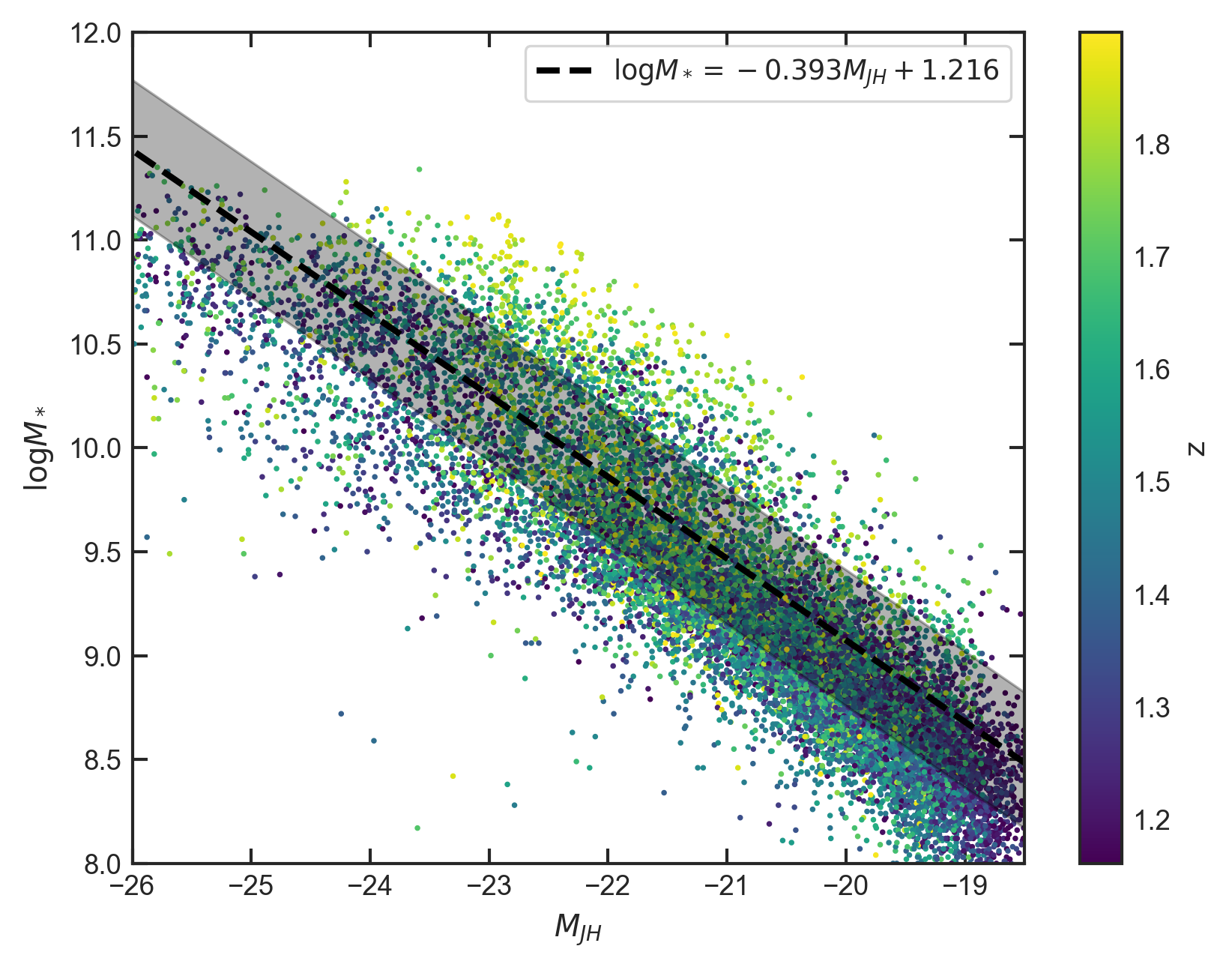}}
    \caption{Stellar mass vs.~absolute JH magnitude for our photo-$z$ sample of $1.2 < z < 1.9$ galaxies with Equation \ref{eq:jhabs} overlaid. The black shaded region contains 95\% of the sample from which Equation \ref{eq:jhabs} was derived. The stellar masses are derived from the \texttt{EAZY} SED-fitting code and use interpolated K-corrections; the linear relation is based on the results from SED fits from \mcsed.  Considering this difference, Equation \ref{eq:jhabs} describes the photometric sample quite well.}
    \label{fig:jhabsphot}
\end{figure}

As a rough test of its universality, Figure~\ref{fig:jhabsphot} illustrates the utility of Equation \ref{eq:jhabs} for our photo-$z$ sample (\S \ref{sec:emvsphot}). For the K-corrections, we used a linear 2-D interpolation on redshift and apparent JH magnitude in the (J+JH+H) band based on the ELG relationships. We used the \texttt{EAZY} results for stellar mass. Considering how different the parameter values from different SED fitting codes can be \citep[e.g.,][]{Leja2019}, the decent agreement between the observations and the ELG-based stellar mass -- absolute magnitude relation (with a slightly different true slope and intercept) is quite impressive and supports the use of Equation~(\ref{eq:jhabs}) for other samples.


To further investigate the role of Equation \ref{eq:jhabs} in the context of galaxy evolution, we downloaded the Granada SDSS catalog \citep{Ahn2014,MonteroDorta2016}, which includes galaxy stellar masses from the Baryon Oscillation Spectroscopic Survey.  These masses use a \citet{Kroupa01} initial mass function, place little-restriction on galaxy age, and include dust in their fits; these assumptions are broadly consistent with those used in our \mcsed (\S \ref{sec:individual}) analysis. The sample of $\sim 1.5$ million galaxies has a median redshift of $0.505$, as compared to our average redshift of $1.463$. Thus, on average, the rest-frame wavelength of the $i$-band in the Granada-SDSS catalog is close to that of the F125W band in our data (without considering K-corrections).

The K-corrections of the SDSS galaxies have been calculated relative to a fiducial redshift of $z_0 = 0.55$.  In contrast, our K-corrections are computed relative to $z_0 = 0.$  To compare the two samples, we therefore we need to apply an additional K-correction to crudely shift the SDSS galaxies to the standard $z_0 = 0$ frame.  Since the rest-frame wavelength of a $z \sim 0.55$ galaxy observed through the SDSS $i$-band (effective wavelength $\lambda_{\rm eff} = 7457.89$), is close to the central wavelength of the SDSS $g$-filter ($\lambda_{\rm eff} = 4671.78$), we can use the $g-r$ colors of the galaxies to inform this additional shift.

\begin{equation}
    M_{I,{z_0=0}} \sim 2*M_{I,{z_0=0.55}} - M_{G,{z_0=0.55}} + 2.5\log 1.55
\end{equation}

\begin{figure}
    \centering
    \resizebox{\hsize}{!}{
    \includegraphics{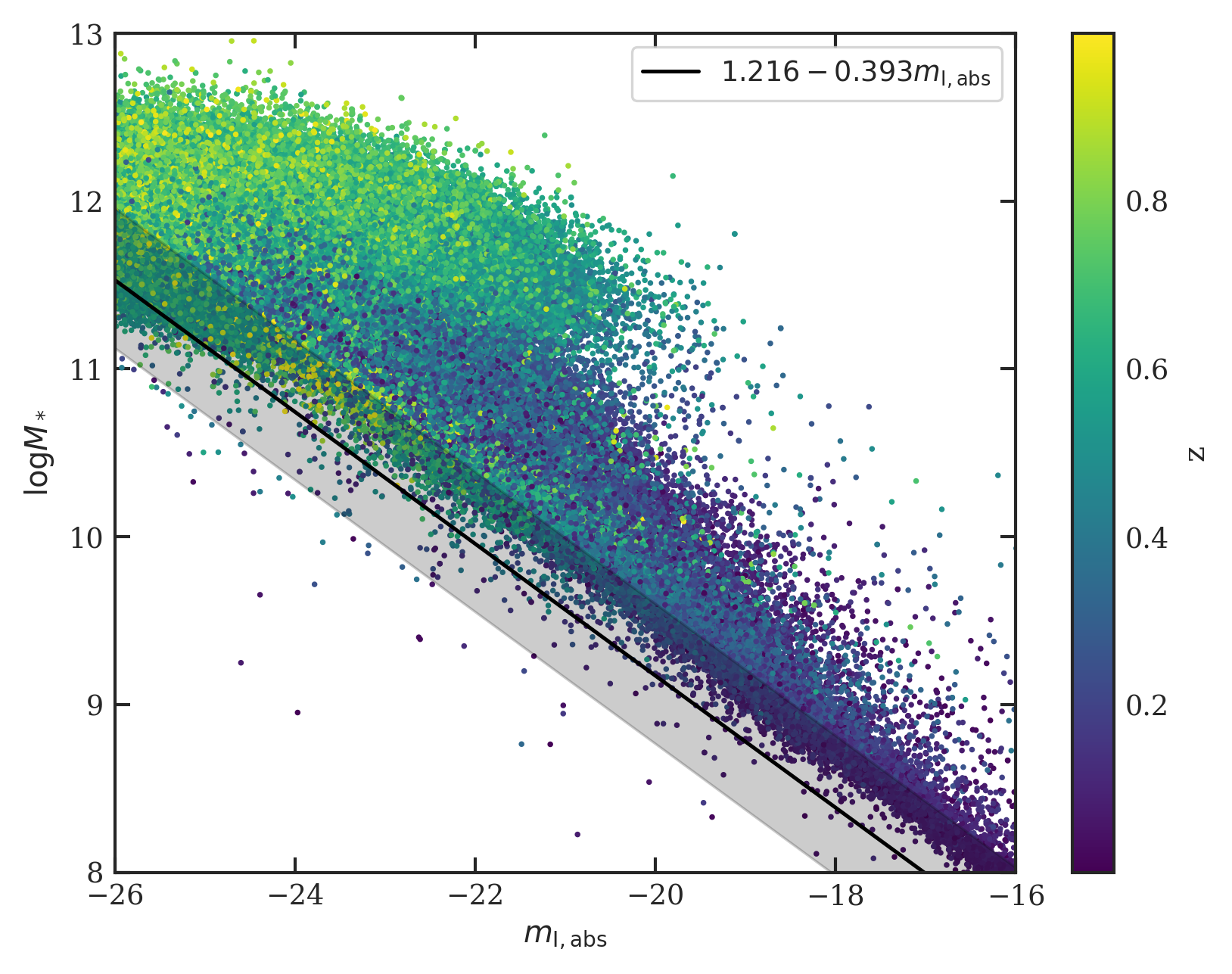}}
    \caption{Stellar mass vs.\ absolute (observed) $i$-band magnitude in the SDSS Granada catalog. The black line is Equation \ref{eq:jhabs}, which applies to our 3D-HST sample, but with JH magnitude converted into $i$-band magnitude using the average ratio of the rest-frame spectrum at the effective wavelengths of the two filters. The black shaded region contains 95\% of the sample from which Equation \ref{eq:jhabs} was derived and accounts for variance in the spectrum ratios.}
    \label{fig:sdss}
\end{figure}

Figure \ref{fig:sdss} shows the stellar mass vs.~absolute $i$-band magnitude for galaxies in the $z \sim 0.55$ SDSS sample. The black line is Equation \ref{eq:jhabs} (stellar mass vs.~absolute JH magnitude) with JH magnitudes replaced by $i$-band magnitudes. 

The conversion from JH magnitude to $i$-band magnitude is based on the average flux-density ratio between a $z \sim 1.5$ ELG's \mcsed spectrum at rest-frame wavelengths 13927 \AA~(the effective wavelength of the F125W + F140W + F160W filter) and 7458 \AA\ (the effective wavelength of the SDSS $i$ band). The standard deviation of these ratios, along with the natural spread we observe in Equation \ref{eq:jhabs}, is taken into account in the black shaded region of the figure.

The reason we can use such a simple conversion and still achieve a decent comparison between the two samples is apparent in Figure \ref{fig:stackspec}: the optical and NIR spectra of ELGs are very well behaved and strongly correlated. Furthermore, the very gradual SED slope ensures that the flux at 1.4 \um is just slightly larger than that at 0.75 \um. 


At low redshift, there is a band of galaxies that extends down to the faintest sources that is approximately parallel to the relation we find at $1.2<z<1.9$.  This suggests that the red/NIR SEDs of the SDSS galaxies are similar to those of our higher redshift ELGs.  But the $z \sim 0.55$ stellar mass-magnitude relation has a different normalization (higher by $\sim 0.3 - 0.6$ dex), implying that high-$z$ sources are more luminous at the same stellar mass. This phenomenon is consistent with the observed evolution of the star-forming main-sequence, as at higher redshift, the SFR is generally higher for a fixed stellar mass \citep[e.g.,][]{Speagle2014}.

Of course, since the SDSS galaxies were defined using a magnitude-limited sample of objects, we cannot discount the possibility that the different normalization is caused by incompleteness;
the bottom end of the stellar mass vs.~magnitude relation may have an artificial cutoff imposed upon it. 


Finally, the SDSS data have a much larger spread of points at the brighter end of the mass function.  This is most likely due to the inclusion of massive, quiescent galaxies that have SED shapes that differ substantially from those of the ELGs. As shown above, the simple relation between absolute magnitude and stellar mass does not apply to these objects.

\section{Conclusion}\label{sec:disc}

Stellar mass is a crucial quantity for constraining hydrodynamical and semi-analytic models of galaxy formation and evolution. There have been many studies measuring stellar masses at high redshift, but due to the method of sample selection, these have typically been limited to high-mass galaxies. The 3D-HST Treasury program \citep[GO-11600, 12177, 12328;][]{Brammer2012,Momcheva2016} is a survey with the 1.08-1.67 \um G141 grism on \textit{HST}'s WFC3 to target rest-frame optical emission lines beyond $z \gtrsim 1$. Emission-line galaxies have been shown to have a broad range of properties, and span a wide range of stellar mass.  The 3D-HST program serves as a trailblazer for future grism surveys such as \textit{Euclid} and \textit{RST}. 

The 3D-HST team provides grism redshifts and emission line strengths for nearly 80,000 sources down to $\jhmag = 26$ \citep{Momcheva2016}. The data products greatly benefit from the wealth of information available in the CANDELS fields \citep{Grogin2011,Koekemoer2011}, which are coincident with the survey area of the 3D-HST observations. \cite{Skelton2014} carefully incorporated 147 public data sets into a comprehensive photometric catalog from 0.3-8 \um for almost all 3D-HST sources.  Here, we have augmented the data set with deep NUV photometry from \textit{GALEX} and the UVOT instrument on \textit{Swift}. 

To create a clean sample of ELGs, we followed (and slightly modified) the procedure motivated by \cite{Zeimann2014} and fully described by \cite{Bowman2019}: we pre-selected those 3D-HST $\jhmag < 26$ between $1.2<z<1.9$ with well-restricted redshift estimates (68\% confidence intervals $\Delta z < 0.05$). We then manually assigned a set of five flags to every source. Objects made it to the ELG sample if they had clear emission lines with little doubt about line identity, little to no contamination unaccounted for by the 3D-HST pipeline, and reasonable SEDs given the \cite{Skelton2014} photometry.

To account for AGN activity, we cross-correlated our galaxy sample with the results of deep \textit{Chandra} surveys. We found 72 AGN in our ELG sample of 4350 objects: an AGN fraction of $1.66\%$ (with a fraction of $2.44\%$ in GOODS-S, which has the deepest observations).  After removing the AGN, we performed a stacking analysis on the remaining objects; the results of this procedure suggest that the vast majority of our galaxies are normal star-forming objects. However, it is possible that Compton-thick AGN are lurking within the sample. We used the \cite{Donley2012} IRAC ($3.6-8.0$ \um observed-frame) criteria to help identify highly obscured sources, finding 39 sources not identified via X-rays. While this is a good first step, it is still difficult to identify low-luminosity or host-dominated obscured AGN. While future studies with more extensive MIR data should be able to help identify AGN that cannot be distinguishable from their X-ray or NIR emission, such objects do no affect the conclusions of this paper.


We ran all objects in the ELG sample through the \mcsed SED-fitting code, in order to determine the galaxies' physical properties, including stellar mass, SFR, and internal extinction. \mcsed generally yields reasonable SED fits for our objects, although some parameters are poorly constrained.  Metallicity, which was fit as a free parameter, is poorly constrained by the stellar SED, but the estimates are vastly improved with the inclusion of emission line ratios from the 3D-HST catalog.


We find that stellar mass is correlated with stellar metallicity. Given that given that measurements of stellar metallicity in star forming galaxies are dominated by light from young stars, fitted stellar metallicities may actually be more representative of gas-phase metallicities considering the young stars were born in nearly-present-day ISM. Therefore, our data an be said to be consistent with the mass-metallicity relation found by \cite{Erb2006}, albeit with a large amount of scatter.

Our most important result is given by Equation \ref{eq:jhabs} and Figure \ref{fig:massvsjhabs}: we find that the single parameter most closely linked to stellar mass is the absolute near-infrared magnitude, as measured by the F125W, F140W, F160W, or the combination of all three filters.  At the redshifts under consideration, these filters broadly cover the rest-frame optical region of the spectrum (before the application of a K-correction). In other words, for sources between redshifts $1.2 \lesssim z \lesssim 1.9$ one can use just a single photometric measurement and a redshift value to get a good estimate of stellar mass, one of the most important physical quantities for a galaxy. 

Equation \ref{eq:jhabs} will be particularly useful for examining the products of the upcoming \textit{Euclid} and \textit{RST} missions, which will generate millions of galaxies similar to those in the 3D-HST catalog. Equation \ref{eq:jhabs} provides a computationally simple but quite accurate and mostly model-independent way of determining stellar masses for star-forming galaxies with stellar masses $11.0 \gtrsim \log{M_*/\msun}\gtrsim 8.5$ and absolute NIR magnitudes $-26 \lesssim M\lesssim -19$.

By investigating the Granada SDSS catalog \citep{Ahn2014,MonteroDorta2016}, we find that a similar relationship persists at much lower redshifts (median $z \sim 0.5$) with the SDSS $i$-band (effective wavelength $7458$ \AA), although the normalization of the distribution is different: at higher redshift, sources with the same mass are more luminous. A more thorough analysis that takes into account incompleteness in both samples would make a strong impact on observational studies of galaxy evolution.

The vast amounts of multiwavelength data in the CANDELS/3D-HST fields allows for a host of related studies. For example, using a statistically valid method to evaluate gas-phase metallicities with multiple strong-line indicators \citep{GrasshornGebhardt2016}, we can investigate the Fundamental Metallicity Relation in this redshift range and probe whether galaxy morphological parameters factor into the correlation. With several large-volume surveys in progress or soon to begin, we can look forward to detailed studies of the effects of environment as well.

\acknowledgments

We thank Brian Lemaux for a meticulous and detailed referee report that really helped us refine the direction of the paper and address any contentious statements made in the initial submission.

We thank Joel Leja for insightful conversations regarding our results. G.N.~thanks Alex Belles for helping him with obtaining \textit{Swift} data and having useful discussions in general.

This work has made use of the Rainbow Cosmological Surveys Database, which is operated by the Centro de Astrobiolog{\'i}a (CAB/INTA), partnered with the University of California Observatories at Santa Cruz (UCO/Lick,UCSC). This work is based on observations taken by the CANDELS Multi-Cycle Treasury Program with the NASA/ESA HST, which is operated by the Association of Universities for Research in Astronomy, Inc., under NASA contract NAS5-26555. This research has made use of NASA’s Astrophysics Data System. This research has made use of the SVO Filter Profile Service (http://svo2.cab.inta-csic.es/theory/fps/) supported from the Spanish MINECO through grant AYA2017-84089. Computations for this research were performed on the Pennsylvania State University’s Institute for Computational and Data Sciences’ Roar supercomputer.  The Institute for Gravitation and the Cosmos is supported by the Eberly College of Science and the Office of the Senior Vice President for Research at the Pennsylvania State University.

This material is based upon work supported by the National Science Foundation Graduate Research Fellowship under Grant No. DGE1255832. Any opinion, findings, and conclusions or recommendations expressed in this material are those of the authors(s) and do not necessarily reflect the views of the National Science Foundation.

\nocite{Rodrigo2012,Rodrigo2020}

%

\vspace{5mm}
\facilities{HST (WFC3), Spitzer (MIPS), Herschel (PACS, SPIRE), GALEX, Swift(UVOT)}


\software{AstroPy \citep{Astropy2013,Astropy2018}, SciPy \citep{Scipy2001,Scipy2020}, CLOUDY \citep{Ferland1998,Ferland2013}, FSPS \citep{Conroy2009,Conroy2010SPSM}, \mcsed \citep{Bowman2020}, R-CRAN \citep{R} }




\bibliography{sample63}{}
\bibliographystyle{aasjournal_mod}



\end{document}